%% file: ANA-HDBS-2019-02-PAPER.tex
\pdfoutput=1
\documentclass[cernpreprint, PAPER, atlasdraft=false, UKenglish, texmf, orcidlogo]{atlasdoc}
 
\usepackage[abbrevref=false]{atlaspackage}
\usepackage{graphicx}
\usepackage{multirow}
\usepackage{bm}
\usepackage{atlasbiblatex}
 
\usepackage{atlasphysics}

\graphicspath{{logos/}{figures/}}
 
\usepackage{ANA-HDBS-2019-02-defs}

\AtlasTitle{Search for non-resonant Higgs boson pair production in the $2b + 2\ell + E_\mathrm{T}^\mathrm{miss}$ final state in $pp$ collisions at $\sqrt{s} = \SI{13}{TeV}$ with the ATLAS detector}
 
\AtlasAbstract{
A search for non-resonant Higgs boson pair ($HH$) production is presented, in which one of the Higgs
bosons decays to a $b$-quark pair ($b\bar{b}$)
and the other decays to $WW^*$, $ZZ^*$, or $\tau^+\tau^-$, with in each case a final state with
$\ell^+\ell^- +$ neutrinos ($\ell = e, \mu$).
The analysis targets separately the gluon-gluon fusion and vector boson fusion production modes.
Data recorded by the ATLAS detector in proton-proton collisions at a centre-of-mass energy of \SI{13}{TeV}
at the Large Hadron Collider, corresponding to an integrated luminosity of \SI{140}{fb^{-1}}, are used in this analysis.
Events are selected to have exactly two $b$-tagged jets and two leptons with opposite electric charge
and missing transverse momentum in the final state.
These events are classified using multivariate
analysis algorithms to separate the $HH$ events from other Standard Model processes.
No evidence of the signal is found. The observed (expected) upper limit on the
cross-section for non-resonant Higgs boson pair production is determined to be 9.7 (16.2) times the
Standard Model prediction at \SI{95}{\percent} confidence level.
The Higgs boson self-interaction coupling parameter $\kappa_\lambda$ and
the quadrilinear coupling parameter $\kappa_{2V}$ are each separately constrained by this analysis to
be within the ranges ${[-6.2, 13.3]}$ and ${[-0.17, 2.4]}$, respectively, at \SI{95}{\percent}
confidence level, when all other parameters are fixed.
}
 
\author{The ATLAS Collaboration}
 
\AtlasRefCode{HDBS-2019-02}
 
\PreprintIdNumber{CERN-EP-2023-233}

\arXivId{2310.11286}

\AtlasJournal{JHEP}
\AtlasJournalRef{JHEP 02 (2024) 037}
\AtlasDOI{10.1007/JHEP02(2024)037}

\hypersetup{pdftitle={ATLAS document},pdfauthor={The ATLAS Collaboration}}
 
\begin{document}
 
\maketitle
 
\tableofcontents

\newpage
\section{Introduction}

The Standard Model (SM) of particle physics employs the Higgs mechanism~\cite{Englert:1964et, Higgs:1964ia,
Higgs:1964pj, Guralnik:1964eu, Higgs:1966ev, Kibble:1967sv} to formulate a theoretical model
in which the weak gauge bosons ($W, Z$) and the fermions acquire mass.
Following the discovery of the Higgs boson, referred to as $H$, by the ATLAS and CMS collaborations \cite{HIGG-2012-27, CMS-HIG-12-028}
at the Large Hadron Collider (LHC) in 2012, one of the remaining open questions is about the structure of the
Higgs potential which is still largely unconstrained.
The rate of the double Higgs production ($HH$) process is sensitive to the trilinear
Higgs self-interaction coupling $\lambda_3$, i.\.e. the coupling
describing a vertex with three Higgs bosons interacting with one another.
It is common practice to define a Higgs boson trilinear coupling
modifier as the actual trilinear coupling value divided by its SM expectation value: $\kappa_\lambda
\equiv \lambda_3/\lambda_3^{SM}$.
 
\begin{figure}[b]
\centering
\subfloat[]{
\includegraphics[width=0.4\textwidth]{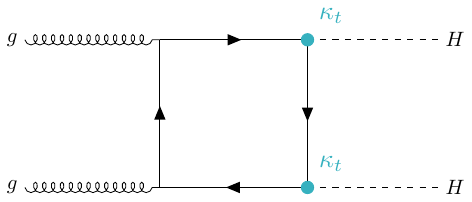}
\label{fig:hh_diagram_ggF_box}
}
\subfloat[]{
\includegraphics[width=0.4\textwidth]{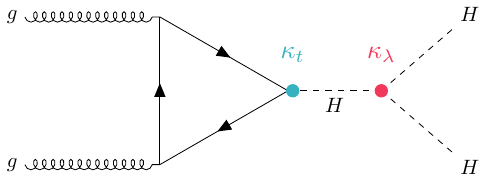}
\label{fig:hh_diagram_ggF_tri}
}
\caption{The leading order (LO) Feynman diagrams for the gluon-gluon fusion process of Higgs boson pair production at
the LHC, where $\kappa_\lambda$ denotes the Higgs boson trilinear coupling modifier
$\kappa_\lambda \equiv \lambda_3/\lambda_3^{SM}$ and $\kappa_t$ the coupling modifier
relative to the top-quark-Higgs-boson coupling.
}
\end{figure}

The dominant $HH$ production process at the LHC is through gluon-gluon fusion (ggF), which involves
either a top-quark box-loop (also referred to as a \textit{box-diagram})
or decay of an off-shell Higgs boson (also referred to as a \textit{triangle-diagram}) at lowest order; the Feynman diagrams are shown in
\cref{fig:hh_diagram_ggF_box,fig:hh_diagram_ggF_tri} respectively.
The \textit{box-diagram} and \textit{triangle-diagram} interfere destructively, leading to a small
cross-section for the $gg\rightarrow HH $ processes,
$\sigma_{\mathrm{ggF}} = \num{31.1}^{+\SI{6.7}{\percent}}_{-\SI{23.2}{\percent}}\,\si{fb}$, calculated at the next-to-next-to-leading
order (NNLO) and including finite top-quark mass effects for a Higgs boson mass of $m_H=\SI{125}{GeV}$
\cite{Dawson:1998py,Borowka:2016ehy,Baglio:2018lrj,deFlorian:2013jea,Shao:2013bz,deFlorian:2015moa,Grazzini:2018bsd,Baglio:2020wgt}.
 
\begin{figure}
\centering
\subfloat[]{\includegraphics[width=0.32\textwidth]{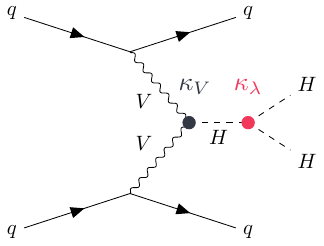}}
\subfloat[]{\includegraphics[width=0.32\textwidth]{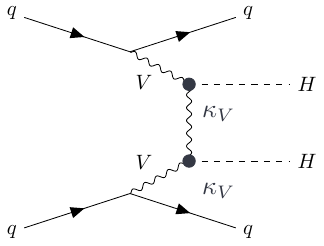}}
\subfloat[]{\includegraphics[width=0.32\textwidth]{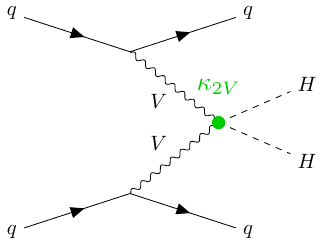}}
\caption{The LO Feynman diagrams for the vector-boson-fusion process of Higgs boson pair production, where
$\kappa_\lambda$, $\kappa_V$, and $\kappa_{2V}$
are the coupling modifiers related to the $HHH$, $HVV$, and $HHVV$ vertices.
}
\label{fig:hh_diagram_VBF}
\end{figure}

The vector-boson-fusion (VBF) process is the sub-leading Higgs boson pair production mode, with a
cross-section of $\sigma_\mathrm{VBF}=
\num{1.726}\pm\SI{2.1}{\percent}\,\si{fb}$ calculated at
next-to-next-to-next-to-leading order (N$^3$LO) QCD for a Higgs boson
mass of $m_H=\SI{125}{GeV}$ \cite{Dreyer:2018qbw}. The production of $HH$ via fusion of vector bosons
($V$) is not only sensitive to
$\kappa_\lambda$ but also depends on the coupling modifier of the Higgs boson to vector
bosons ($\kappa_V$) and on another coupling modifier related to the quartic vertex involving two vector bosons and two Higgs bosons
($\kappa_{2V}$). The corresponding diagrams are shown in \cref{fig:hh_diagram_VBF}.
 
In the presence of physics beyond the Standard Model (BSM), the $HH$ production cross-section can be altered by modifying
the value of the self-coupling $\lambda_3$, leading to values for $\kl$ different from the SM prediction
as discussed in Refs. \cite{DiLuzio:2017tfn,Kribs:2017znd}. Similarly, modifications of the coupling to
vector bosons can alter $\kvv$.
Probing the Higgs boson self-coupling can therefore provide additional information about the validity of the SM.
 
Searches for SM $HH$ production through the decay channels
$HH\rightarrow bb\gamma\gamma$ \cite{HDBS-2018-34}, $HH \rightarrow 4b$ \cite{HDBS-2018-41} and $HH \rightarrow bb\tau\tau$ \cite{HDBS-2018-40}
were performed using the full Run~2 dataset collected by the ATLAS experiment at
$\sqrt{s} = \SI{13}{TeV}$. The results from these three channels have been combined \cite{HDBS-2022-03}
to improve the search sensitivity.
The observed (expected in the absence of $HH$ production) combined upper limit on the production rate
of SM Higgs boson pairs, at the \SI{95}{\percent} confidence level (CL), was found to be $2.4$ ($2.9$) times the SM prediction.
The value of $\kappa_\lambda$ was
constrained to be within the observed (expected) range $[-0.6, 6.6]$ ($[-2.1, 7.8]$) at \SI{95}{\percent}
CL and the value for $\kvv$ was constrained to be within the observed (expected) range $[0.1, 2.0]$
($[0.0, 2.1]$) at \SI{95}{\percent} CL \cite{HDBS-2022-03}.
With searches for $HH$ production trough the decay channels $HH \rightarrow 4b$
\cite{CMS-HIG-20-005,CMS-B2G-22-003}, $HH \rightarrow bb\tau\tau$ \cite{CMS-HIG-20-010}, $HH \rightarrow
WW^*/\tau\tau + WW^* / \tau\tau$ \cite{CMS-HIG-21-002}, $HH \rightarrow bb\gamma\gamma$
\cite{CMS-HIG-19-018} and $HH \rightarrow 4\ell + bb$ \cite{CMS-HIG-20-004},
similar results have also been obtained by the CMS experiment; according to those results, the observed
(expected) \SI{95}{\percent} upper limit on the production rate of SM Higgs boson pairs is $3.4$ ($2.5$)
times the SM prediction,
$\kappa_\lambda$ is constrained to be within the observed range of $[-1.2, 6.5]$
at \SI{95}{\percent} CL \cite{CMS-HIG-22-001} and
$\kvv$ is constrained to be within the observed range of $[-0.67, 1.4]$
at \SI{95}{\percent} CL \cite{CMS-HIG-22-001}.
 
This paper focuses on a data analysis using \bbllmet ($\ell = e, \mu$) final states arising from
different $HH$ decay channels,
\begin{equation*}
HH\rightarrow b\bar{b} + WW^* / ZZ^* / \tau^+\tau^- \rightarrow b\bar{b} + \ell^+\ell^- + \text{neutrinos},
\end{equation*}
with one of the Higgs bosons decaying to a $b$-quark pair ($b\bar{b}$) and the other decaying to a boson
($WW^*$, $ZZ^*$) or a $\tau^+\tau^-$ pair, which further decays to ${\ell^+\ell^- + \text{neutrinos}}$.
The leptons can be of different flavour.
The experimental signature is characterised by two $b$-tagged jets with the invariant mass $m_{bb}$ close to $m_H$, two leptons with opposite charges, and large missing transverse energy $E_\mathrm{T}^\mathrm{miss}$ due to neutrinos escaping detection.
The decay channel with a Higgs boson decaying to two $Z$ bosons and the subsequent
final state with two leptons and two quark-initiated jets from the $ZZ$ system, ${bbZZ \rightarrow
bb 2\ell 2q}$, is also considered. Although this final state does not have $E_\mathrm{T}^\mathrm{miss}$
due to the presence of neutrinos,
it can have $E_\mathrm{T}^\mathrm{miss}$ due to detector and reconstruction inefficiencies.
Dominant background arises from top-quark processes ($t\bar{t}$, $Wt$ and $t\bar{t}V$ with $V=W,Z$), and
production of a $Z$ boson associated with heavy-flavour jets ($Z$+HF).
The search uses the full Run~2 dataset collected with the ATLAS experiment
in proton-proton ($pp$) collisions at \SI{13}{TeV}, corresponding to a total integrated luminosity of \SI{140}{\ifb}.
 
A search for $HH$ production in the \bbllmet final state considering only $HH \rightarrow bbWW^*$ using
the full Run~2 dataset has already been performed~\cite{HDBS-2018-33};
the work presented in this paper significantly improves the analysis techniques
by optimizing a deep neural network (DNN) to better classify the \bbllmet
events into $HH$ signal or background and considers additional decay channels, leading to a factor of two better sensitivity.
Furthermore, studies of $HH$ production through the VBF process are included for the first time in the \bbllmet final state.
In addition, the compatibility of the data with non-SM values of the \kl and $\kappa_{2V}$ parameters is constrained in this paper.


\section{ATLAS detector}

 
\newcommand{\AtlasCoordFootnote}{
ATLAS uses a right-handed coordinate system with its origin at the nominal interaction point (IP)
in the centre of the detector and the \(z\)-axis along the beam pipe.
The \(x\)-axis points from the IP to the centre of the LHC ring,
and the \(y\)-axis points upwards.
Cylindrical coordinates \((r,\phi)\) are used in the transverse plane,
\(\phi\) being the azimuthal angle around the \(z\)-axis.
The pseudorapidity is defined in terms of the polar angle \(\theta\) as \(\eta = -\ln \tan(\theta/2)\).
Angular distance is measured in units of \(\Delta R \equiv \sqrt{(\Delta\eta)^{2} + (\Delta\phi)^{2}}\).}
 
The ATLAS detector~\cite{PERF-2007-01} at the LHC covers nearly the entire solid angle around the collision point.\footnote{\AtlasCoordFootnote}
It consists of an inner tracking detector surrounded by a thin superconducting solenoid, electromagnetic and hadron calorimeters,
and a muon spectrometer incorporating three large superconducting air-core toroidal magnets.
 
The inner-detector system (ID) is immersed in a \SI{2}{\tesla} axial magnetic field
and provides charged-particle tracking in the range \(|\eta| < 2.5\).
The high-granularity silicon pixel detector covers the vertex region and typically provides four measurements per track,
the first hit normally being in the insertable B-layer (IBL) installed before Run~2~\cite{ATLAS-TDR-19,PIX-2018-001}.
It is followed by the silicon microstrip tracker (SCT), which usually provides eight measurements per track.
These silicon detectors are complemented by the transition radiation tracker (TRT),
which enables radially extended track reconstruction up to \(|\eta| = 2.0\).
The TRT also provides electron identification information
based on the fraction of hits (typically 30 in total) above a higher energy-deposit threshold corresponding to transition radiation.
 
The calorimeter system covers the pseudorapidity range \(|\eta| < 4.9\).
Within the region \(|\eta|< 3.2\), elec\-tro\-magnetic calorimetry is provided by barrel and
endcap high-granularity lead/liquid-argon (LAr) calorimeters,
with an additional thin LAr presampler covering \(|\eta| < 1.8\)
to correct for energy loss in material upstream of the calorimeters.
Hadron calorimetry is provided by the steel/scintillator-tile calorimeter,
segmented into three barrel structures within \(|\eta| < 1.7\), and two copper/LAr hadron endcap calorimeters.
The solid angle coverage is completed with forward copper/LAr and tungsten/LAr calorimeter modules
optimised for electromagnetic and hadronic energy measurements respectively.
 
The muon spectrometer (MS) comprises separate trigger and
high-precision tracking chambers measuring the deflection of muons in a magnetic field generated by the superconducting air-core toroidal magnets.
The field integral of the toroids ranges between \num{2.0} and \SI{6.0}{\tesla\metre}
across most of the detector.
Three layers of precision chambers, each consisting of layers of monitored drift tubes, cover the region \(|\eta| < 2.7\),
complemented by cathode-strip chambers in the forward region, where the background is highest.
The muon trigger system covers the range \(|\eta| < 2.4\) with resistive-plate chambers in the barrel and thin-gap chambers in the endcap regions.
 
Interesting events are selected by the first-level trigger system implemented in custom hardware,
followed by selections made by algorithms implemented in software in the high-level trigger~\cite{TRIG-2016-01}.
The first-level trigger accepts events from the \SI{40}{\MHz} bunch crossings at a rate below \SI{100}{\kHz},
which the high-level trigger further reduces in order to record events to disk at about \SI{1}{\kHz}.
 
An extensive software suite~\cite{ATL-SOFT-PUB-2021-001} is used in data simulation, in the reconstruction
and analysis of real and simulated data, in detector operations, and in the trigger and data acquisition
systems of the experiment.


\section{Data and Monte Carlo samples}

The analysis is based on the full Run~2 dataset collected in $pp$ collisions at the LHC with a centre-of-mass energy of $\sqrt{s} = \SI{13}{\TeV}$ and a \SI{25}{\ns} bunch crossing interval.
The data used in this
analysis are required to have been recorded while all relevant components of the
ATLAS detector were in their nominal operating conditions \cite{DAPR-2018-01}.
The integrated luminosity of the dataset collected over the full Run~2 data-taking period and suitable
for physics analysis corresponds to \SI{140}{\ifb} \cite{DAPR-2021-01}.
The candidate events with oppositely charged leptons are selected based on a combination of single-lepton
and di-lepton triggers.\footnote{Distinct sets of single-lepton triggers are used for electrons and muons.
Di-lepton triggers require either two electrons, two muons, or one electron and one muon.} The use of a
given trigger depends on the flavour and the transverse momenta
($\pt$) of the two leptons in the event, and on the data-taking period. Single-lepton triggers
have $\pt$ thresholds between \SI{21}{GeV} and \SI{28}{GeV}. Di-lepton triggers are only considered
if no single-lepton trigger criteria are met and have $\pt$ thresholds as
low as $13 (9)\,\si{GeV}$ for the leading (sub-leading) lepton.
 
Monte Carlo (MC) simulation~\cite{SOFT-2010-01} is generally used to model the $HH$ signal and SM background processes.
The generation of the physics processes of interest, including the underlying event and immediate decays, are carried out by dedicated event generators.
The generated events are then passed through a simulation of the ATLAS detector based on \GEANT~\cite{Agostinelli:2002hh}.
Some background MC event samples, including $Z$+jets with $\SI{10}{GeV} < m_{\ell\ell} < \SI{40}{GeV}$ and the $t\bar{t}$ and $Wt$
alternative samples used for estimating systematic uncertainties, are simulated through the ATLAS fast simulation framework, Atlfast-II
(AF2)~\cite{SOFT-2010-01} instead. Additionally, the effect of multiple interactions in the same and neighbouring bunch
crossings (pile-up) is modelled by overlaying the simulated hard-scattering event with
inelastic $pp$ events generated with \PYTHIA[8.186]~\cite{Sjostrand:2007gs}
using the \NNPDF[2.3lo] set of parton distribution functions (PDF)~\cite{Ball:2012cx} and the
A3 set of tuned parameters~\cite{ATL-PHYS-PUB-2016-017}.
A summary of the event samples used for the simulation of the signal and background processes is shown in \cref{tab:mc_bkgd_nominal}.
 
\begin{table}
\caption{Summary of nominal SM background processes considered in the analysis along with a
description of the event generators used for matrix element (ME) generation, the set of parton
distribution functions (PDF), the hadronisation, parton shower (PS) and underlying event (UE) model, and the
underlying event tune.}
\centering
\resizebox{\textwidth}{!}{
\begin{tabular}{lcccccc}
\toprule
Process                          & ME Generator                  & ME PDF          & PS/UE model                   & UE Tune         \\
\midrule
SM $HH$ (ggF)                    & \POWHEGBOX[v2]                & \PDFforLHC[15nlo] & \Pythia[8.244]                & A14             \\
SM $HH$ (VBF)                    & \MGNLO[2.7.3]                 & \NNPDF[3.0nlo]  & \Pythia[8.244]                & A14             \\
\midrule
$\ttbar$                         & \POWHEGBOX[v2]                & \NNPDF[3.0nlo]  & \Pythia[8.230]                & A14             \\
Single-top                       & \POWHEGBOX[v2]                & \NNPDF[3.0nlo]  & \Pythia[8.230]                & A14             \\
$\ttbar+W/Z$                     & \MGNLO[2.3.3]                 & \NNPDF[3.0nlo]  & \Pythia[8.210]                & A14             \\
\midrule
$\ensuremath{W/Z+\mathrm{jets}}$ & \Sherpa[2.2.1]                & \NNPDF[3.0nnlo] & \Sherpa[2.2.1]                & \Sherpa default \\
\midrule
$WW, WZ, ZZ$                     & \Sherpa[2.2.1]/\Sherpa[2.2.2] & \NNPDF[3.0nnlo] & \Sherpa[2.2.1]/\Sherpa[2.2.2] & \Sherpa default \\
\midrule
$ggF, H$                         & \POWHEGBOX[v2]                & \NNPDF[3.0nlo]  & \Pythia[8.212]                & AZNLO           \\
$VBF, H$                         & \POWHEGBOX[v2]                & \NNPDF[3.0nlo]  & \Pythia[8.230]                & AZNLO           \\
$WH, ZH$                         & \POWHEGBOX[v2]                & \NNPDF[3.0nlo]  & \Pythia[8.230]/\Pythia[8.186] & AZNLO           \\
$t\bar{t}H$                      & \POWHEGBOX[v2]                & \NNPDF[3.0nlo]  & \Pythia[8.230]                & A14             \\
\bottomrule
\end{tabular}
}
\label{tab:mc_bkgd_nominal}
\end{table}
 
SM $HH$ signal events produced via the ggF production mechanism are generated with \POWHEGBOX[v2] at next-to-leading order
(NLO) interfaced with \Pythia[8.244]~\cite{Sjostrand:2014zea} using the \PDFforLHC[15nlo] \cite{Butterworth:2015oua}
PDF set and the A14 tune~\cite{ATL-PHYS-PUB-2014-021}.
For parton shower variations, \Herwig[7.1.5] \cite{HERWIGBellm:2019zci} is used instead. Samples for non-SM
ggF production, i.\,e. with $\kl \neq 1$ are obtained from simulated samples at different
values of these coupling modifiers and combined using morphing techniques \cite{HDBS-2018-58}
with detailed validation studies of this procedure available in Ref.~\cite{ATL-PHYS-PUB-2019-007}.
The SM VBF $HH$ signal samples are generated with \MGNLO[2.7.3] \cite{Alwall:2014hca}
using the \NNPDF[3.0nlo]~\cite{Ball:2014uwa} PDF set and the A14 tune~\cite{ATL-PHYS-PUB-2014-021}
interfaced to \Pythia[8.244]~\cite{Sjostrand:2014zea} with \Herwig[7.2.1]~\cite{HERWIGBellm:2019zci} used for
parton shower variations. As is the case in Ref.~\cite{HDBS-2022-03}, signal templates with coupling modifiers ${(\kl \neq 1,
\kvv \neq 1)}$ are obtained by linear combination of six samples with different values for the $\kl$ and
$\kvv$ parameters \cite{Bishara:2016kjn}. Therefore five more independent samples with ${(0, 0)}$, ${(1,
0.5)}$, ${(1, 3)}$, ${(2, 1)}$ and ${(10, 1)}$ are generated as well with \MADGRAPH interfaced to \Pythia{8}. To
vary only $\kvv$, three independent samples are enough: two more independent samples with $\kvv = 1.5$ and
$\kvv = 2$ are generated in the same manner. The decays of bottom and charm hadrons of all SM $HH$ signal
events were performed by \EVTGEN[1.6.0]~\cite{Lange:2001uf}.
 
SM background processes are generally estimated
using simulation, although the cross-sections of the dominant background
components, $t\bar{t}$, $Wt$ and $Z+$jets, are constrained using  control regions (CRs) to better
predict their contribution in the signal regions (SRs).
The estimate is therefore said to be \textit{semi-data-driven} with the normalization being constrained
from data and the shape being taken from simulation.
For processes not constrained from dedicated CRs, both the shape and the overall normalization are taken from simulation. In addition, the fake-lepton background is estimated in a data-driven way.
 
SM top-quark production ($t\bar{t}$) and the production of top-quarks in association with $W$ bosons ($Wt$) contribute as a significant background contamination in the \bbllmet final state.
At NLO, non-trivial interference arises between these two processes that may be enhanced in phase-space regions with high fractions of $Wt$ events \cite{TOPQ-2017-05}.
The two most commonly used schemes to remove the overlap between these two processes are the diagram removal (DR) and diagram subtraction (DS) schemes \cite{Frixione:2008yi}.
The former is used in the present analysis to remove the overlapping events while the latter is used to evaluate the systematic uncertainty in corresponding background event yields.
 
The decays of bottom and charm hadrons of all SM background events not simulated with \Sherpa were
performed by \EVTGEN[1.6.0]~\cite{Lange:2001uf}.
 
Production of \ttbar events is modelled using the
\POWHEGBOX[v2]~\cite{Frixione:2007nw,Nason:2004rx,Frixione:2007vw,Alioli:2010xd}
generator at NLO with the \NNPDF[3.0nlo]~\cite{Ball:2014uwa} PDF set
and the \hdamp parameter\footnote{The
\hdamp parameter is a re-summation damping factor and one of the
parameters that controls the matching of \POWHEG matrix elements to
the parton shower and thus effectively regulates the
high-\pT radiation against which the \ttbar system recoils.} set
to 1.5 times the top-quark mass \cite{ATL-PHYS-PUB-2016-020}.  The events are interfaced
to \PYTHIA[8.230]~\cite{Sjostrand:2014zea} to model the parton shower,
hadronisation, and underlying event, with parameters set according
to the A14 tune~\cite{ATL-PHYS-PUB-2014-021} and using the \NNPDF[2.3lo]
set of PDFs~\cite{Ball:2012cx}.
In order to reduce mis-modelling, events are reweighted
\cite{serkinTreatmentTopquarkBackgrounds2021} to match the predictions at NNLO QCD and NLO EW based on
the true \pt-value of the top-quark (i.\,e. not the anti-top-quark) provided by the MC event samples.
 
Production of $Wt$ events is modelled by the
\POWHEGBOX[v2]~\cite{Re:2010bp,Nason:2004rx,Frixione:2007vw,Alioli:2010xd}
generator at NLO in QCD using the five-flavour scheme and the
\NNPDF[3.0nlo] set of PDFs~\cite{Ball:2014uwa}.
The events are interfaced to \PYTHIA[8.230]~\cite{Sjostrand:2014zea} using the A14
tune~\cite{ATL-PHYS-PUB-2014-021} and the \NNPDF[2.3lo] set of
PDFs~\cite{Ball:2012cx}.
 
The $V+$jets samples ($V=W$ or $Z$) are simulated with \Sherpa[2.2.1] \cite{Bothmann:2019yzt} interfaced
with \NNPDF[3.0nlo] \cite{Ball:2012cx} for both the matrix element (ME) calculation and the parton
shower (PS) tuning.  The merging of different parton multiplicities is achieved through a matching
scheme based on the CKKW-L \cite{Lonnblad:2001iq, Lavesson:2005xu} merging technique using a scale
parameter of $Q_{cut} = \SI{20}{\GeV}$.  The modelling of higher jet multiplicities relies on the parton
shower algorithm. The parton shower and underlying event models used are the ones provided internally by
\Sherpa.  The \Sherpa[2.2.1] generator adopts a full 5-flavour scheme, with massless $b$- and $c$-quarks
in the matrix elements, while massive quarks can be produced in the parton shower.  The $V+$jets samples
are split according to the $\pT$ of the vector boson and the scalar sum of jet transverse momenta,
$H_\mathrm{T}$, of the event, introducing a cut at generation level and producing samples for different
slices in max($H_\mathrm{T},p_\mathrm{T}^V$) where $p_\mathrm{T}^V$ is defined as the transverse momentum
of the true lepton pair from the decay of the $V$ boson.  The samples are also generated by applying
different filters to select the flavour composition of the jets produced in association with the $V$
boson.
 
Di-boson processes with four charged leptons, three charged leptons and one neutrino, or two charged leptons and two neutrinos are
simulated with the \Sherpa[2.2.2] event generator \cite{Bothmann:2019yzt}, and di-boson processes with
leptons and jets are simulated with \Sherpa[2.2.1] \cite{Bothmann:2019yzt}. The matrix elements contain all
diagrams with four electroweak vertices. They are calculated for up to one parton at \NLO and up to three
partons at LO using Comix \cite{Gleisberg:2008fv} and OpenLoops \cite{Cascioli:2011va} and merged with
the \Sherpa parton shower according to the ME+PS@NLO prescription \cite{Hoeche:2012yf}.
The NNPDF{3.0nnlo} PDF set is used in conjunction with dedicated parton shower tuning developed by the \Sherpa authors. The event generator cross-sections are used in this case (already at NLO).
 
Finally, single Higgs boson samples from all the main production modes ggF, VBF, Higgs-strahlung ($WH$,
$ZH$) and associated production with a pair of top-quarks ($t\bar{t}H$) are considered. For all Higgs
boson samples the normalization accounts for the decay branching ratio calculated
with \textsc{HDECAY}~\cite{Djouadi:1997yw,Spira:1997dg,Djouadi:2006bz}
and \PROPHECY~\cite{Bredenstein:2006ha,Bredenstein:2006rh,Bredenstein:2006nk}.
Higgs boson production via gluon--gluon fusion is simulated at NNLO accuracy in
QCD using \POWHEGBOX[v2]~\cite{Hamilton:2013fea,Hamilton:2015nsa,Alioli:2010xd,Nason:2004rx,Frixione:2007vw}.
The simulation achieves NNLO accuracy for arbitrary inclusive \(gg\to H\) observables by reweighting the Higgs boson
rapidity spectrum in \textsc{Hj}-\MINLO~\cite{Hamilton:2012np,Campbell:2012am,Hamilton:2012rf} to that of HNNLO~\cite{Catani:2007vq}.
The \NNPDF[3.0nlo]~\cite{Ball:2014uwa} PDF set and the \AZNLO{} tune~\cite{STDM-2012-23}
of \PYTHIA[8.212]~\cite{Sjostrand:2014zea} are used.
The gluon--gluon fusion prediction from the MC event samples is normalised to the
N$^3$LO cross-section in QCD plus electroweak (EW) corrections
at NLO~\cite{deFlorian:2016spz,Anastasiou:2016cez,Anastasiou:2015ema,Dulat:2018rbf,Harlander:2009mq,Harlander:2009bw,Harlander:2009my,Pak:2009dg,Actis:2008ug,Actis:2008ts,Bonetti:2018ukf,Bonetti:2018ukf}.
Higgs boson production via vector-boson fusion is simulated with
\POWHEGBOX[v2]~\cite{Nason:2009ai,Alioli:2010xd,Nason:2004rx,Frixione:2007vw}
and interfaced with \PYTHIA[8.230]~\cite{Sjostrand:2014zea} for parton shower and non-perturbative effects,
with parameters set according to the \AZNLO{} tune~\cite{STDM-2012-23}.
The \POWHEGBOX prediction is accurate to NLO and uses
the \NNPDF[3.0nlo]~\cite{Ball:2014uwa} PDF set.
It is normalised to an approximate-NNLO QCD cross-section
with NLO EW corrections~\cite{Ciccolini:2007jr,Ciccolini:2007ec,Bolzoni:2010xr}.
Higgs boson production in association with a vector boson is simulated using
\POWHEGBOX[v2]~\cite{Nason:2009ai,Alioli:2010xd,Nason:2004rx,Frixione:2007vw} and interfaced with \PYTHIA[8.230]~\cite{Sjostrand:2014zea} for
parton shower and non-perturbative effects. The \POWHEGBOX{} prediction is accurate to NLO for \(VH\) boson plus one-jet production.
The loop-induced \(gg\to ZH\) process is generated separately at leading order. The \NNPDF[3.0nlo]~\cite{Ball:2014uwa} PDF set
and the \AZNLO{} tune~\cite{STDM-2012-23} of \PYTHIA[8.186]~\cite{Sjostrand:2014zea} are used. The MC
prediction is normalised to cross-sections calculated at NNLO in QCD with NLO EW corrections for \(q\bar{q}/qg \to VH\) and at NLO
and next-to-leading-logarithm accuracy in QCD for
\(gg \to ZH\)~\cite{Ciccolini:2003jy,Brein:2003wg,Brein:2011vx,Altenkamp:2012sx,Denner:2014cla,Brein:2012ne,Harlander:2014wda}.
The production of ttH events is modelled using the
\POWHEGBOX[v2]~\cite{Frixione:2007nw,Nason:2004rx,Frixione:2007vw,Alioli:2010xd,Hartanto:2015uka}
generator at NLO with the \NNPDF[3.0nlo]~\cite{Ball:2014uwa} PDF set.
The events are interfaced to \PYTHIA[8.230]~\cite{Sjostrand:2014zea}~using
the A14 tune~\cite{ATL-PHYS-PUB-2014-021} and the
\NNPDF[2.3lo]~\cite{Ball:2014uwa} PDF set.


\section{Object definition and event selection}
\label{sec:selection}

Proton-proton interaction vertices are reconstructed in events with at least two tracks, each with ${\pt > \SI{0.5}{\GeV}}$.
The primary hard-scatter vertex for each event is defined as the one with the highest value of the sum of squared track transverse momenta~\cite{ATL-PHYS-PUB-2015-026}.
 
Electron candidates are reconstructed from energy deposits measured in the electromagnetic calorimeter which are
matched to ID tracks \cite{EGAM-2021-01}. They are required to satisfy $|\eta| < 2.47$, excluding the calorimeter transition
region $1.37 < |\eta| < 1.52$, and have a transverse momentum $\pt > \SI{10}{GeV}$. Electron candidates are required to satisfy
a \enquote{medium} identification criterion based on the use of shower shape, track–cluster matching and TRT
parameters in a likelihood-based algorithm \cite{EGAM-2021-01}.
Additionally, a \enquote{loose} isolation requirement \cite{EGAM-2021-01} is applied to electron
candidates to ensure that they are well separated from other objects in the event.
This requirement is based on the momentum of nearby tracks and calorimeter energy deposits within a cone
around the electron candidate.
 
Muon candidates are reconstructed from high-quality tracks found
in the MS \cite{PERF-2015-10}. A matching of these tracks to ID tracks is required in the region $|\eta|
< 2.5$. Muon candidates are
required to have $|\eta| < 2.7$ and $\pt > \SI{9}{GeV}$, and to satisfy a \enquote{medium} identification criterion \cite{MUON-2018-03}.
Additionally, a \enquote{loose} isolation requirement \cite{MUON-2018-03} is imposed on muon candidates to ensure that they are well separated from other objects in the event.
This requirement is based on investigating the nearby activities within a cone around the muon candidate.
 
Jets are reconstructed using the anti-$k_t$ algorithm \cite{Cacciari:2008gp, Fastjet} with a radius
parameter of $R = 0.4$. It is applied to $|\eta| < 4.5$ noise-suppressed positive-energy
topological energy clusters \cite{PERF-2014-07,ATL-PHYS-PUB-2015-036} and charged particle tracks
processed using a particle-flow algorithm \cite{PERF-2015-09}. Pile-up is taken into account in the
formation of topological energy clusters \cite{PERF-2016-04}. Jet candidates are required to have $\pT >
\SI{20}{GeV}$ and those with $|\eta| < 2.5$ are considered to
select the candidate jets for the reconstruction of the hadronic decay of the Higgs to $b$-tagged jets
while jets with $|\eta| > 2.5$ are considered \enquote{forward} jets.
To reject jet candidates originating from pile-up interactions, they must satisfy a tight pile-up
suppression requirement based on a multivariate method \cite{ATLAS-CONF-2015-029}.
The method removes jets that appear to be inconsistent with the primary vertex and have $\pt
< \SI{60}{\GeV}$ \cite{ATL-PHYS-PUB-2015-026}.
In the end, a neural network-based $b$-tagging algorithm \cite{FTAG-2019-07} is used to identify jets
containing $b$-hadrons ($b$-tagged jets) at a \SI{77}{\percent} efficiency working point in simulated \ttbar
events. The rejection rate for this working point is $\sim\!6$ for jets originating from $c$-quarks and
$\sim\!200$ for jets originating from light quarks \cite{FTAG-2019-07}.
 
The missing transverse momentum, $\boldsymbol{p}_\mathrm{T}^{\text{miss}}$, the magnitude of which is denoted as \MET, quantifies the amount of energy and momentum carried by invisible particles that do not interact with the detector.
It is determined as the magnitude of the negative vectorial sum of the transverse momenta of the selected
and calibrated physics objects, including jets, electrons, and muons, and inner detector tracks
from the hard-scatter collision vertex not associated with any physics object \cite{PERF-2016-07}.
 
Different reconstructed objects can share the same detector signature, leading to ambiguities in the
identification of these objects. To resolve these ambiguities, as in \cite{HDBS-2018-40}, an overlap
removal procedure is performed sequentially, such that only objects that survive the previous step are
considered in the following steps. The steps are as follows: if any electrons share a track, only the
highest $\pt$ electron is kept. If a hadronically decaying $\tau$-lepton candidate is
within\footnote{This is the angular distance $\Delta R$ considering the rapidity $y$ instead of the
pseudorapidity $\eta$ and is defined as $\Delta R_y \equiv \sqrt{(\Delta y)^{2} + (\Delta\phi)^{2}}$} $\Delta R_y
= 0.2$ of any electron or muon, it is removed. If an electron and a muon share a
track, the muon is kept only if it is associated with a signature in the muon spectrometer. Any jet within
$\Delta R_y = 0.2$ of an electron and subsequent any electron within $\Delta R_y = 0.4$ of any jet
is removed. Any jet within $\Delta R_y = 0.2$ of a muon, or having an inner detector track ghost-matched
\cite{Cacciari:2007fd} to a muon within $\Delta R_y = 0.2$ of the jet, is removed if it has fewer then
three associated tracks. Any muon within $\Delta R_y = 0.4$ of a jet is removed as well as any jet within
$\Delta R_y = 0.2$ of a hadronically decaying $\tau$-lepton candidate.

The analysis selects candidate events that contain exactly two opposite-charge light leptons, either
electrons or muons, and exactly two $b$-tagged jets. While the signal processes contain sources of
$\MET$,
an explicit selection on $\MET$ is not performed, to ensure high training statistics available for the MVA
discriminants.
To suppress the contribution of misidentified jets, the analysis excludes events that contain jets satisfying one of the bad jet criteria \cite{ATLAS-CONF-2015-029}.
The yields of the ggF and VBF signal samples at different preselection steps are summarized in
\cref{tab:selection:cutflow}; about an order of magnitude of reduction in the event yields is observed
from this initial selection.
 
\begin{table}
\begin{center}
\caption{
Cutflow for event selection using SM $gg/qq \rightarrow HH $ signal samples in various decay channels.
For both ggF and VBF signal samples, the SM $HH$ cross-section, $\sigma$, and branching ratio,
$\mathcal{B}$, are assumed when computing event yields for a luminosity of $\mathcal{L} =
\SI{140}{\ifb}$. Efficiencies are different for $bbZZ(\to{}2\ell{}2\nu)$ compared to
$bbZZ(\to{}2\ell{}2q)$ since the initial number of events considers $Z\rightarrow\tau\tau$
while the former does not.
}
\label{tab:selection:cutflow}
\resizebox{\columnwidth}{!}{
\begin{tabular}{@{}lcccccccc@{}}
\toprule
\multirow{2}{*}{ggF and VBF event selection cut}
&\multicolumn{2}{c}{$bbWW$}      &\multicolumn{2}{c}{ $bb\tau\tau$}
&\multicolumn{2}{c}{$bbZZ(\to{}2\ell{}2\nu)$} &\multicolumn{2}{c}{$bbZZ(\to{}2\ell{}2q)$} \\
 
& ggF & VBF& ggF & VBF& ggF & VBF& ggF & VBF \\
\midrule
Initial number of events ($\mathcal{L} \times \sigma \times \mathcal{B}$)                             &70& 3.9   &39& 2.2    &3.8& 0.21             &18& 1.0            \\
$N_\text{leptons} = 2$, opposite sign, pass trigger requirement                                       &22& 0.99  &8.3& 0.35   &1.3& 0.057            &3.6& 0.17           \\
$N_{b\text{-jets}} = 2$    &9.8& 0.39  &3.7& 0.14   &0.57& 0.022           &1.6 & 0.067         \\
\bottomrule
\end{tabular}
}
\end{center}
\end{table}
 
\begin{figure}
\centering
\subfloat[][]{
\label{fig:selection:region:sf}
\includegraphics[width=0.48\textwidth]{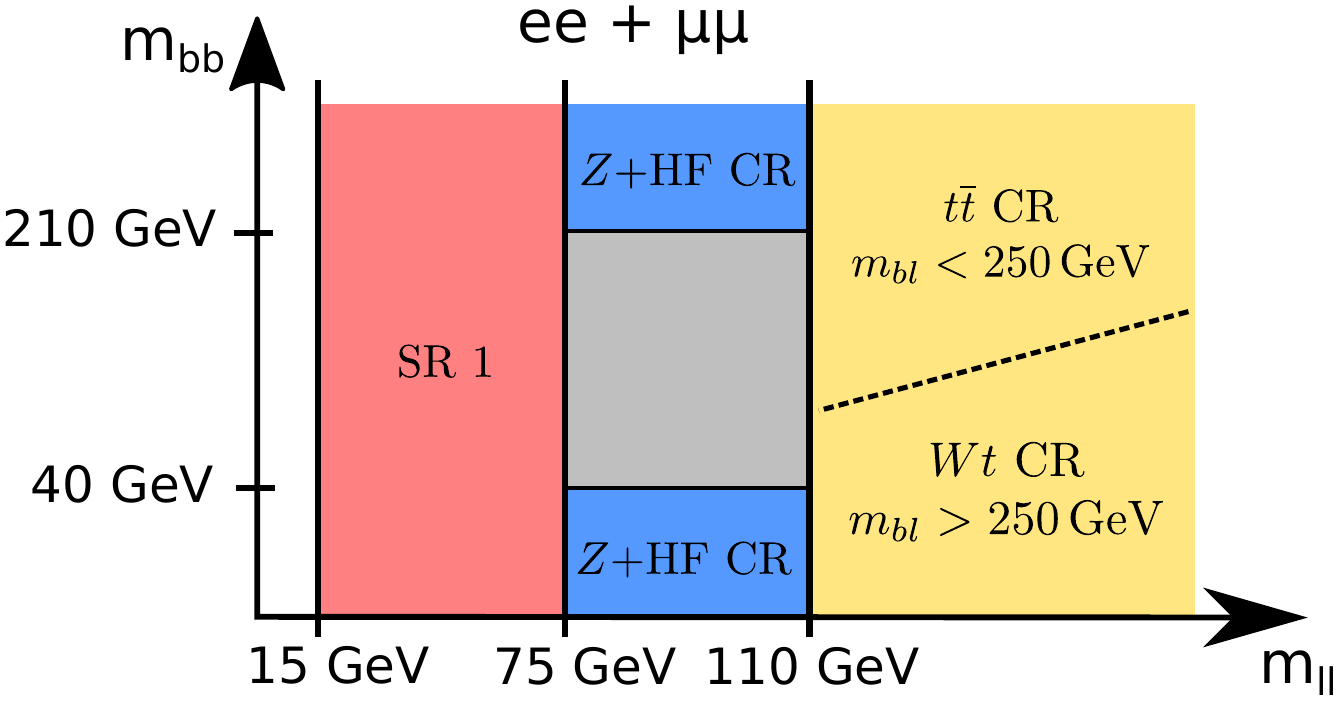}
}
\subfloat[][]{
\label{fig:selection:region:df}
\includegraphics[width=0.48\textwidth]{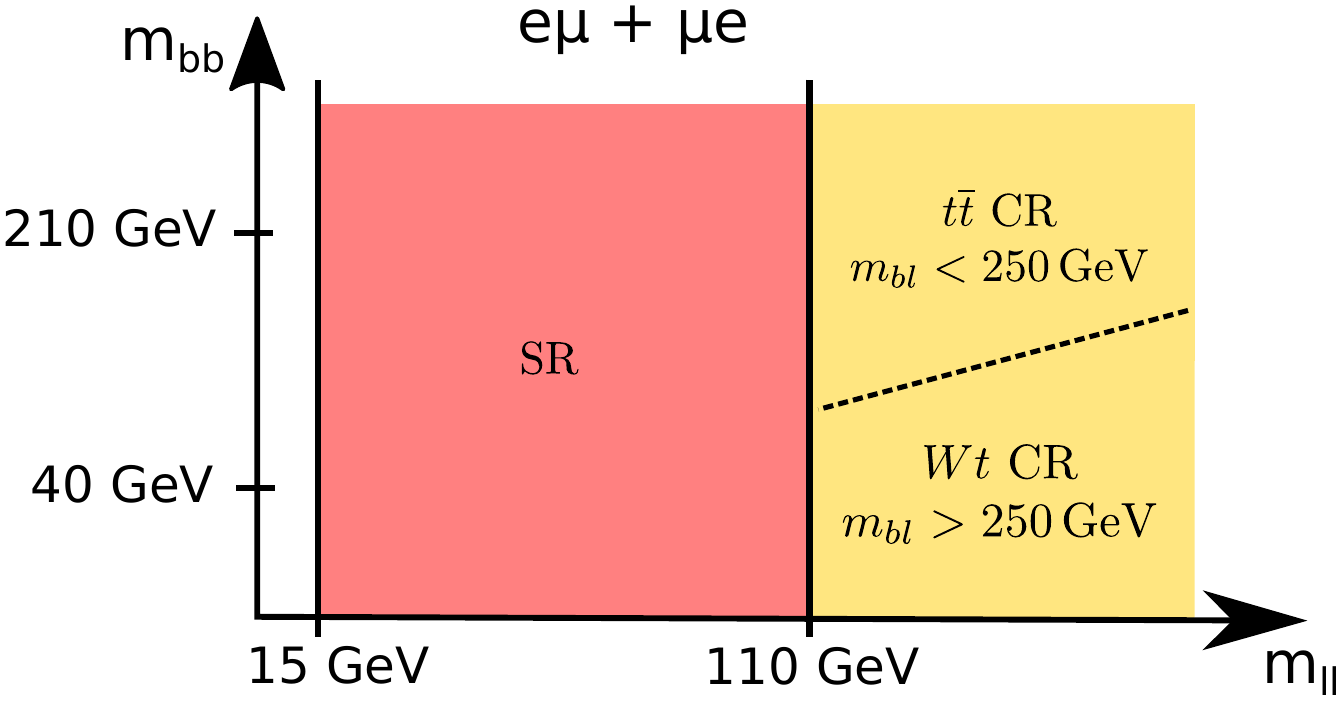}
}
\caption{
Definition of signal and control regions for same lepton flavour (a) and different lepton flavour (b)
events, where either the electron ($e\mu$) is leading in $\pT$ or the muon ($\mu e$). The greyed-out region is excluded as it makes a negligible contribution to the
final results. The $m_{b\ell}$ discriminant variable \cite{TOPQ-2017-05}
is used to further separate the top CR into separate \ttbar and $Wt$ control regions.
}
\label{fig:selection:region}
\end{figure}

The events that satisfy the preselection criteria are further divided into two categories: the signal
region (SR) and dedicated background control regions (CRs) for $Z$ plus heavy flavour jets ($Z$+HF),
$\ttbar$ and $Wt$, as depicted in
\cref{fig:selection:region}. The greyed out region is not considered since it does not add significantly
to the sensitivity. The CRs are used to constrain the background normalization in the SRs. The
SR and CRs are separated based on the invariant mass of the two leptons, with a lower threshold set at
\SI{15}{GeV} and an upper threshold set at \SI{75}{GeV} for same flavour (SF) leptons and \SI{110}{GeV}
for different flavour (DF) leptons, as illustrated in \cref{fig:selection:region:sf} and
\cref{fig:selection:region:df}. The mass of the di-lepton system is required to be $m_{\ell\ell} >
\SI{110}{GeV}$ for the CRs enriched in events coming from top-quark processes. This region is further
split into dedicated \ttbar- and \wt-enriched control regions by requiring\footnote{See
\cref{tab:mva:input_features} for its definition.} $m_{b\ell}$
\cite{TOPQ-2017-05} to be below or above $\SI{250}{\GeV}$, respectively. The $Z$+HF CR is required to
have two SF leptons in the events, and to satisfy the di-lepton invariant mass requirement,
$\SI{75}{\GeV} < m_{\ell\ell} < \SI{110}{\GeV}$, excluding events where the invariant mass of the two
$b$-tagged jets satisfies $\SI{40}{\GeV} < m_{bb} < \SI{210}{\GeV}$.
 
A VBF selection is applied to further categorize the events into two orthogonal ggF/VBF enriched regions
by reverting (for ggF-like regions) or applying (for VBF-like regions) the VBF selection. The VBF selection
requires the presence of at
least two forward jets: these jets must have $\pt > \SI{30}{\GeV}$, at least one pair of these extra
jets must have a minimum pseudorapidity separation larger than \num{4.0}, and at least one pair of these
jets must have a mass larger than $\SI{600}{\GeV}$. The jets pairs considered are not required to be the
same for the
pseudorapidity separation and mass requirements; in less than $\sim\!\SI{1}{\percent}$ of the MC events,
different pairs of jets are chosen. The VBF selection has a relative selection efficiency of
$\sim\!\SI{60}{\percent}$ on VBF signal events. \cref{tab:selection:yield} shows the yields for SM
background processes and non-resonant SM ggF and VBF signals in the SRs and CRs with their statistical
uncertainty taken either directly from MC or from the statistics of the template, derived as described in
\cref{sec:bkg_estimate}, in case for the fake-lepton background. For two minor backgrounds negative
yields are observed due to NLO MC statistical fluctuations; the total is always positive. The SRs are
then used to
extract the final results, with the backgrounds constrained by the CRs, as described later. To enhance
the sensitivity to the signal process and maximize the rejection of the expected SM backgrounds, a
multivariate approach is used to select signal events, as described in the next section.
 
\begin{table}
\begin{center}
\caption{
Pre-fit yields for SM background processes and non-resonant SM ggF and VBF signals in the SRs and CRs.
The event yields in the row \enquote{Fakes} come from the data-driven fake-lepton background
estimate described
in \cref{sec:bkg_estimate}. The event yields for $Z$+jets processes are split into ones with heavy-flavour
(HF) and light-flavour (LF) jets as defined in \cref{sec:bkg_estimate}. A dash represents no
contribution of the respective MC event sample in the respective region. Uncertainties are
from MC statistics and template statistics only.
}
\label{tab:selection:yield}
\resizebox{\columnwidth}{!}{
\begin{tabular}{l *{5}{r@{\,}c@{\,}l}}
\toprule
\textbf{Process}         & \multicolumn{3}{c}{\textbf{ggF-SR}} &
\multicolumn{3}{c}{\textbf{VBF-SR}}    & \multicolumn{3}{c}{\textbf{$t\bar{t}$-CR}}
& \multicolumn{3}{c}{\textbf{$Wt$-CR}}
& \multicolumn{3}{c}{\textbf{$Z$+HF-CR}}        \\
\midrule \midrule
\multicolumn{16}{c}{SM background}                                                                                                                   \\
\midrule
$t\bar{t}$          & 561220  &$\pm$&  150     &  52670  &$\pm$&  50       &  436840  &$\pm$&  130         &  2270  &$\pm$&  10              &  34700  &$\pm$&  40        \\
$t\bar{t}+V$        & 1121  &$\pm$&  4         &  194.7  &$\pm$&  1.9      &  1133  &$\pm$&  5             &  97.0  &$\pm$&  1.1             &  440.1  &$\pm$&  1.9       \\
Single top ($Wt$)        & 16260  &$\pm$&  50       &  1165  &$\pm$&  12        &  14100  &$\pm$&  40           &  2901  &$\pm$&  20              &  1237  &$\pm$&  13         \\
Single top (s/t-channel) & 12.7  &$\pm$&  0.8       &  2.48  &$\pm$&  0.35      &  1.21  &$\pm$&  0.28          &  0.35  &$\pm$&  0.14            &  0.25  &$\pm$&  0.11       \\
$Z\to \ell\ell$ (HF)      & 16090  &$\pm$&  180      &  1178  &$\pm$&  34        &  3610  &$\pm$&  70            &  525  &$\pm$&  11               &  43390  &$\pm$&  260       \\
$Z\to \ell\ell$ (LF)      & 2720  &$\pm$&  170       &  260  &$\pm$&  40         &  600  &$\pm$&  90             &  55  &$\pm$&  8                 &  5470  &$\pm$&  190        \\
$Z\to\tau\tau$ (HF)      & 2200  &$\pm$&  40        &  154  &$\pm$&  13         &  3  &$\pm$&  7                &  1.9  &$\pm$&  0.5              &  4  &$\pm$&  6             \\
$Z\to\tau\tau$ (LF)      & 370  &$\pm$&  50         &  24  &$\pm$&  4           &  -1.3  &$\pm$&  1.5           &  0.11  &$\pm$&  0.06            &  0.8  &$\pm$&  0.5         \\
$W$+jets                   & 0.7  &$\pm$&  0.5        &  0.09  &$\pm$&  0.08      &  -0.2   &$\pm$&  0.4           &  \multicolumn{3}{c}{---}              &  \multicolumn{3}{c}{---}         \\
Diboson                  & 288  &$\pm$&  4          &  32.6  &$\pm$&  0.8       &  159.0  &$\pm$&  2.8          &  39.0  &$\pm$&  0.9             &  226.8  &$\pm$&  3.3       \\
Single Higgs                    & 601.0  &$\pm$&  1.1      &  105.1  &$\pm$&  0.4      &  336.5  &$\pm$&  0.5          &  22.06  &$\pm$&  0.12           &  48.28  &$\pm$&  0.29      \\
Fakes               & 18510  &$\pm$&  170      &  2390  &$\pm$&  60        &  10020  &$\pm$&  140          &  529  &$\pm$&  35               &  1360  &$\pm$&  50         \\
\midrule
Total SM bkg.            & 619390  &$\pm$&  350     & 58170  &$\pm$&  100      & 466810  &$\pm$&  230         & 6440  &$\pm$&  40              & 86890  &$\pm$&  330       \\
\midrule \midrule
\multicolumn{16}{c}{$HH$ signal, ggF}                                                                                                        \\
\midrule
ggF $HH\to{}bbWW$        & 8.318  &$\pm$&  0.016    &  0.857  &$\pm$&  0.005    &  0.00113  &$\pm$&  0.00019    &  0.00033  &$\pm$&  0.00010      &  0.0014  &$\pm$&  0.0002 \\
ggF $HH\to{}bb\tau\tau$  & 3.138  &$\pm$&  0.009    &  0.3284  &$\pm$&  0.0029  &  0.00332  &$\pm$&  0.00029    &  0.00068  &$\pm$&  0.00015      &  0.0047  &$\pm$&  0.0004   \\
ggF $HH\to{}bbZZ$        & 0.633  &$\pm$&  0.005    &  0.0873  &$\pm$&  0.0018  &  0.00083  &$\pm$&  0.00018    &  0.00020  &$\pm$&  0.00009      &  0.0442  &$\pm$&  0.0013   \\
 
\midrule
$\sum$ ggF $HH$          & 12.088  &$\pm$&  0.019   & 1.272  &$\pm$&  0.006    & 0.0053  &$\pm$&  0.0004      & 0.00121  &$\pm$&  0.00020      & 0.0504  &$\pm$&  0.0014   \\
\midrule \midrule
\multicolumn{16}{c}{$HH$ signal, VBF}                                                                                                        \\
\midrule
VBF $HH\to{}bbWW$        & 0.1518  &$\pm$&  0.0014  &  0.2138  &$\pm$&  0.0017  &  0.00013  &$\pm$&  0.00004    &  \multicolumn{3}{c}{---}              &  0.00009  &$\pm$&  0.00004 \\
VBF $HH\to{}bb\tau\tau$  & 0.0537  &$\pm$&  0.0006  &  0.0769  &$\pm$&  0.0007  &  0.000086  &$\pm$&  0.000022  &  0.000048  &$\pm$&  0.000018    &  0.00024  &$\pm$&  0.00004 \\
VBF $HH\to{}bbZZ$        & 0.0097  &$\pm$&  0.0004  &  0.0184  &$\pm$&  0.0006  &  0.000040  &$\pm$&  0.000024  &  0.0000029  &$\pm$&  0.0000016  &  0.00236  &$\pm$&  0.00023 \\
 
\midrule
$\sum$ VBF $HH$          & 0.2152  &$\pm$&  0.0016  & 0.3091  &$\pm$&  0.0019  & 0.00026  &$\pm$&  0.00005    & 0.000051  &$\pm$&  0.000018    & 0.00269  &$\pm$&  0.00024 \\
\midrule \midrule
\multicolumn{16}{c}{$HH$ signal, ggF+VBF}                                                                                                    \\
\midrule
$\sum$ ggF+VBF $HH$      & 12.303  &$\pm$&  0.019   & 1.582  &$\pm$&  0.006    & 0.0055  &$\pm$&  0.0004      & 0.00126  &$\pm$&  0.00020      & 0.0531  &$\pm$&  0.0014   \\
\bottomrule
\end{tabular}
}
\end{center}
\end{table}


\section{Multivariate analyses}

Events that pass the selections described in the previous section are then passed through a MVA to
separate rare signal events from the large amount of background events.
A DNN and a Boosted Decision Tree (BDT) are used to classify events into the ggF and VBF categories, respectively.
The outputs of the MVA models are used as discriminants in the statistical analysis discussed in \cref{sec:result}.
 
In the ggF category, the Keras library \cite{chollet2015keras} with Tensorflow
as the backend \cite{tensorflow2015-whitepaper} is used to design a DNN classifier with a multi-output
architecture to optimise the separation between the ggF $HH$ signal, background from $t\bar{t}$ and $Wt$ and all
other background processes simultaneously. In a first stage, \SI{50}{\percent} of the simulated events
are used to optimize the set of input variables and the hyperparameters of the DNN in a five-fold
cross-validation with a train/test split fraction of $\SI{80}{\percent}/\SI{20}{\percent}$. The final
set of input variables is selected based on the permutation feature
importance \cite{breimanRandomForests2001} to keep only the most
important ones; this strategy allows the analysis to not compromise the performance but reduces the complexity of the
model and therefore its goodness given the finite training statistics.
These input variables are listed in \cref{tab:mva:input_features}, with $\MET$ significance being ranked
as the ninth and $\MET$ being ranked as the twelfth
most important input variables with respect to the overall sensitivity. The data are
well described by MC for these variables.
The hyperparameters of the DNN are optimized through an automatic process using the Optuna
\cite{akibaOptunaNextgenerationHyperparameter2019} package, with the figure of merit being the
\SI{95}{\percent} CL upper limit on the signal strength, i.\,e. the ratio of the measured $HH$ production cross section to its SM prediction,
without considering systematic uncertainties, using the $\mathrm{CL}_s$ prescription~\cite{Read:2002hq}.
The final DNN model includes nine fully connected (FC) layers, each with 512 nodes, and is trained at a
learning rate of 0.00011, with ReLU activations \cite{nairRectifiedLinearUnits2010} used for the FC
layers and softmax activations used for the output layer. It is trained with a two-fold cross-validation
strategy with a train/test split fraction of $\SI{50}{\percent}/\SI{50}{\percent}$.
A dropout rate of 0.3 is applied to the FC layers to prevent over-fitting; as a result the loss curve reaches a
plateau for both training and test datasets at the end of the training. The score of the signal output
node is binned so that the statistical uncertainty of the backgrounds is less than \SI{30}{\percent} and
$\sim\!1$ $HH$ signal event from the combined ggF and VBF production modes is expected in each bin.
The seven bins with the highest DNN output score are further considered for the statistical analysis, leading
to a maximum of $\mathcal{O}(10^2)$ background events being present in a bin to be considered as part of the
final signal region.
 
\begin{table}
\centering
\caption{Input features used for the DNN in the ggF category. Indices 0 and 1 refer to $\pt$-leading
and $\pt$-sub-leading objects respectively.}
\label{tab:mva:input_features}
\begin{tabular}{ll}
\toprule
Input feature                                & Description \\
\midrule
same flavour                                 & unity if final state leptons are $ee$ or $\mu\mu$, zero otherwise \\
$\pt^{\ell}$, $\pt^b$                        & transverse momenta of the leptons, $b$-tagged jets \\
$m_{\ell\ell}$, $\pt^{\ell\ell}$             & invariant mass and the transverse momentum of the di-lepton system \\
$m_{bb}$, $\pt^{bb}$                         & invariant mass and the transverse momentum of the $b$-tagged jet pair system \\
$m_\mathrm{T2}^{bb}$                         & stransverse mass of the two $b$-tagged jets~\cite{Lester:1999tx,Lester:2014yga} \\
$\Delta R_{\ell\ell}$, $\Delta R_{bb}$       & $\Delta R$ between the two leptons and two $b$-tagged jets \\
$m_{b\ell}$                                  & $\min\{\max(m_{b_0 \ell_0}, m_{b_1 \ell_1}), \max(m_{b_0 \ell_1}, m_{b_1 \ell_0}) \}$~\cite{TOPQ-2017-05} \\
$\min\Delta R_{b\ell}$	& minimum $\Delta R$ of all $b$-tagged jet and lepton combinations \\
$m_{bb\ell\ell}$                             & invariant mass of the $bb\ell\ell$ system \\
\MET, \MET-sig                            & missing transverse energy and its significance \cite{ATLAS-CONF-2018-038} \\
$m_\mathrm{T}(\ell_0, E_\mathrm{T}^\mathrm{miss})$ & transverse mass of the $\pt$-leading lepton with
respect to $E_\mathrm{T}^\mathrm{miss}$  \\
$\min m_{\mathrm{T},\ell}$               & minimum value of $m_\mathrm{T}(\ell_0, E_\mathrm{T}^\mathrm{miss})$ and $m_\mathrm{T}(\ell_1, E_\mathrm{T}^\mathrm{miss})$ \\
$H_\mathrm{T2}^\mathrm{R}$                       & measure for boostedness\footnotemark of the two Higgs bosons \\
\bottomrule
\end{tabular}
\end{table}
 
In contrast to the ggF event category, due to limited statistics and a tighter phase
space in the VBF category, a DNN approach cannot be fully exploited for such events.
Therefore, a BDT classifier is trained using the adaptive boosting (\texttt{AdaBoost}) method
and the TMVA framework \cite{TMVA} with a cross-validation setup.
Input variables are listed in \cref{tab:mva:BDTinput_features}, with $\MET$ significance being ranked as
the fourteenth
most important input variable with respect to the overall sensitivity. The data are
well described by MC for these variables.
The training parameters are optimized to have 350 trees with a maximum depth of four and a minimum terminal
node size of \SI{2.5}{\percent}. The signal tree provided to the algorithm consists exclusively of VBF
$HH$ events, while the background consists of ggF $HH$ signal events and other SM background events.
The ggF $HH$ events are classified as background since maximising the sensitivity to the VBF $HH$ production
mode is the goal of this MVA discriminant. A two-fold cross-validation was utilized with a train/test split fraction of
$\SI{50}{\percent}/\SI{50}{\percent}$. Each BDT was employed for the half of the dataset it had not been
trained on, ensuring a reliable evaluation of performance across the entire dataset. After careful
examination, it has been decided to train the BDT on a VBF sample with $\kappa_\lambda = 0$, as it showed
the best overall performance across a set of SM and BSM scenarios. The BDT output score is binned into ten
bins taking into account their signal over background ratio, with the most sensitive bin being required to
have a minimum of $\sim\!2$ background events.
The five bins with the highest BDT output score are further considered for the statistical analysis, leading
to a maximum of $\mathcal{O}(10^3)$ background events being present in a bin to be considered as part of the
final signal region.
 
\footnotetext{$H_\mathrm{T2}^\mathrm{R} = \frac{\left|\boldsymbol{E}_\mathrm{T}^\mathrm{miss} + \boldsymbol{p}_\mathrm{T}^{\ell_0} + \boldsymbol{p}_\mathrm{T}^{\ell_1} \right| + \left|\boldsymbol{p}_\mathrm{T}^{b_0} + \boldsymbol{p}_\mathrm{T}^{b_1} \right|}{\left|\boldsymbol{E}_\mathrm{T}^\mathrm{miss}\right| + \left|\boldsymbol{p}_\mathrm{T}^{\ell_0}\right| + \left|\boldsymbol{p}_\mathrm{T}^{\ell_1}\right| + \left|\boldsymbol{p}_\mathrm{T}^{b_0}\right| + \left|\boldsymbol{p}_\mathrm{T}^{b_1}\right|}$}
 
\begin{table}
\centering
\caption{Input features for the BDT algorithm in the VBF category. The usage of $j$ in variable names
indicates that only non $b$-tagged jets being considered. Indices 0 and 1 refer to $\pt$-leading and
$\pt$-sub-leading objects respectively.}
\label{tab:mva:BDTinput_features}
\begin{tabular}{ll}
\toprule
Input feature	& Description \\
\midrule
$\eta_{\ell_0}$ , $\eta_{\ell_1}$, $\phi_{\ell_0}$, $\phi_{\ell_1}$, $p_\mathrm{T}^{\ell_0}$, $p_\mathrm{T}^{\ell_1}$	& $\eta$, $\phi$, $\pt$ of the $\pt$-(sub)leading lepton \\
$\eta_{b_0}$, $\eta_{b_1}$, $\phi_{b_0}$, $\phi_{b_1}$, $p_\mathrm{T}^{b_0}$, $p_\mathrm{T}^{b_1}$                                              & $\eta$, $\phi$, $\pt$ of the $\pt$-(sub)leading $b$-tagged jet \\
$\eta_{j_0}$, $\eta_{j_1}$, $\phi_{j_0}$, $\phi_{j_1}$, $p_\mathrm{T}^{j_0}$, $p_\mathrm{T}^{j_1}$	& $\phi$, $\eta$, $\pt$ of the $\pt$-(sub)leading non $b$-tagged jet \\
$E_\mathrm{T}^\mathrm{miss}$, $\phi^{E_\mathrm{T}^\mathrm{miss}}$, $E_\mathrm{T}^\mathrm{miss}$-sig	& missing transverse energy, its $\phi$ and significance \cite{ATLAS-CONF-2018-038} \\
$\pt^{bb}$, $\Delta R_{bb}$, $\Delta \phi_{bb}$,  $m_{bb}$	& $\pt$, $\Delta R$, $\Delta \phi$ and invariant mass of di-$b$-jet system \\
 
$\pt^{\ell\ell}$, $\Delta R_{\ell\ell}$, $\Delta \phi_{\ell\ell}$, $m_{\ell\ell}$,
$\phi_\mathrm{centrality}^{\ell\ell}$	& $\pt$, $\Delta R$, $\Delta \phi$, $\pt$ and
centrality\footnotemark{} of di-leptons system \\
$\pt^{bb\ell\ell}$, $m_{bb\ell\ell}$	& $\pt$ and invariant mass of the $bb\ell\ell$ system \\
$\pt^{bb\ell\ell + E_\mathrm{T}^\mathrm{miss}}$, $m_{bb\ell\ell + E_\mathrm{T}^\mathrm{miss}}$	& $\pt$ and invariant mass of $bb\ell\ell$ + $E_\mathrm{T}^\mathrm{miss}$ system \\
$m_{\ell\ell + E_\mathrm{T}^\mathrm{miss}}$	& invariant mass of di-lepton + $E_\mathrm{T}^\mathrm{miss}$ system \\
$p_\mathrm{T}^{E_\mathrm{T}^\mathrm{miss} + \ell\ell}$, $\Delta\phi_{\MET,\ell\ell}$	& $p_\mathrm{T}$ of and $\Delta\phi$ between $E_\mathrm{T}^\mathrm{miss}$ and di-lepton system \\
$\pt^\mathrm{tot}$	& $\pt$ of $bb\ell\ell + E_\mathrm{T}^\mathrm{miss} + $ $\pT$-leading and -sub-leading jet \\
$m_\mathrm{tot}$	& invariant mass of $bb\ell\ell + E_\mathrm{T}^\mathrm{miss} + $ $\pT$-leading and -sub-leading jet \\
$m_t^\mathrm{KLF}$	& Kalman fitter top-quark mass \cite{Erdmann:2013rxa} \\
$\min\Delta R_{\ell_{0} j}$, $\min\Delta R_{\ell_{1} j}$	& minimum $\Delta R$ between $\pt$-(sub)leading $\ell\text{-}j$ couples \\
$\sum m_{\ell j}$	& sum of the invariant masses of all $\ell+$jet combinations \\
$\max p^{jj}_{\mathrm{T}}$, $\max m_{jj}$	& maximum $p_\mathrm{T}$ and invariant mass of any two non $b$-tagged jets \\
$\max\Delta\eta_{jj}$, $\max\Delta\phi_{jj}$	& maximum $\Delta \eta$ and $\Delta \phi$ between any two non $b$-tagged jets\\
$\min\Delta R_{b\ell}$	& minimum $\Delta R$ of all $b$-tagged jet and lepton combinations \\
$N_\mathrm{forward\ jets}$, $N_{j}$	& number of forward jets, number of non $b$-tagged jets \\
$m_\mathrm{T2}^{bb}$	& stransverse mass of the two $b$-tagged jets \cite{Lester:1999tx,Lester:2014yga} \\
$m_\mathrm{coll}$	& collinear mass (reconstruction of $m_{\tau \tau}$) \cite{elaginNewMassReconstruction2011} \\
$m_\mathrm{MMC}$	& value of the MMC algorithm (reconstruction of $m_{\tau \tau}$) \cite{elaginNewMassReconstruction2011} \\
\bottomrule
\end{tabular}
\end{table}


\section{Background estimation}
\label{sec:bkg_estimate}

The dominant background processes expected to contribute to the signal region are top quark pair
production (\ttbar), single top-quark in association with a $W$ boson (\wt), and $Z/\gamma^*$ production in
association with heavy-flavour jets. The contributions of these background processes are constrained in
dedicated CRs as defined in \cref{sec:selection}.
 
The simulated $Z$+jets background events are divided into heavy flavour (HF) and light flavour (LF), based
on the generator-level information on the true origin of the jets. If at least one of the two \pt-ordered
leading jets in the $Z+$jets sample is matched at generator level with either a $b$- or $c$-quark, events
are classified as \enquote{heavy flavour}. All other $Z+$jets events, i.\,e. events where at least one of
the two \pt-leading jets is a light flavour jet, are classified as light flavour. As the cross-section of $Z$ boson production
in association with heavy flavour jets is known to be mis-modelled \cite{CMS-EWK-11-015,STDM-2012-15}, the
MC prediction is constrained with data in control regions enriched in the $Z+$HF processes.
 
The contribution from events containing photons or jets that are mis-identified as leptons as well as
leptons from the hadronic decays of heavy flavour quarks is collectively referred as \enquote{fake-lepton} background.
It is estimated using a data-driven method.
 
Firstly, the corresponding CR is defined to have the same selections as the SR, except reverting the opposite-sign (OS) requirement of the di-lepton system in the SR definition to have only same-sign (SS) lepton pairs.
The contribution of fake-lepton events in the SS region is then estimated by subtracting the predicted
contribution from prompt leptons of SM backgrounds from the observed number of the SS di-lepton events in
the data.
Next, transfer factors binned as a function of sub-leading lepton $\pt$ are calculated as the ratio of the number of OS fake-lepton events
to the number of SS fake-lepton events,
\begin{equation*}
f_\mathrm{SS\rightarrow OS} = \frac{N^\mathrm{fake}_\mathrm{MC,OS}}{N^\mathrm{fake}_\mathrm{MC,SS}},
\end{equation*}
as estimated from the background MC events; they
range from $1.2$ to $1.9$. In the end, the number of fake-lepton events in the SRs is extrapolated by
applying the transfer factors to the number of fake-lepton events estimated in the SS region:
\begin{equation*}
N^\mathrm{fake}_\mathrm{OS}  = f_\mathrm{SS\rightarrow OS} \times \left(N_\mathrm{data,SS} -
N^\mathrm{prompt}_\mathrm{MC,SS}\right).
\end{equation*}
 
\footnotetext{The centrality $\phi_\mathrm{centrality}^{\ell\ell}$ indicates the location
of the missing transverse energy with respect to the two final state leptons in the $\phi$-plane. It is
one when the direction of $\MET$ lies directly between the two leptons and $1/\sqrt{2}$ if it is
aligned with one of the leptons.
It is defined as $\phi_\mathrm{centrality}^{\ell\ell} = (1/\sqrt{2}) (\phi_A + \phi_B) / (\phi_A^2 + \phi_B^2)^{1/2}$ with $\phi_A = \sin(\phi_{\MET} - \phi_{\ell_0}) / \sin(\phi_{\ell_1} - \phi_{\ell_0})$ and $\phi_B = \sin(\phi_{\ell_1} - \phi_{\MET}) / \sin(\phi_{\ell_1} - \phi_{\ell_0})$.
}


\section{Systematic uncertainties}

\begin{figure}
\centering
\includegraphics[width=.8\textwidth]{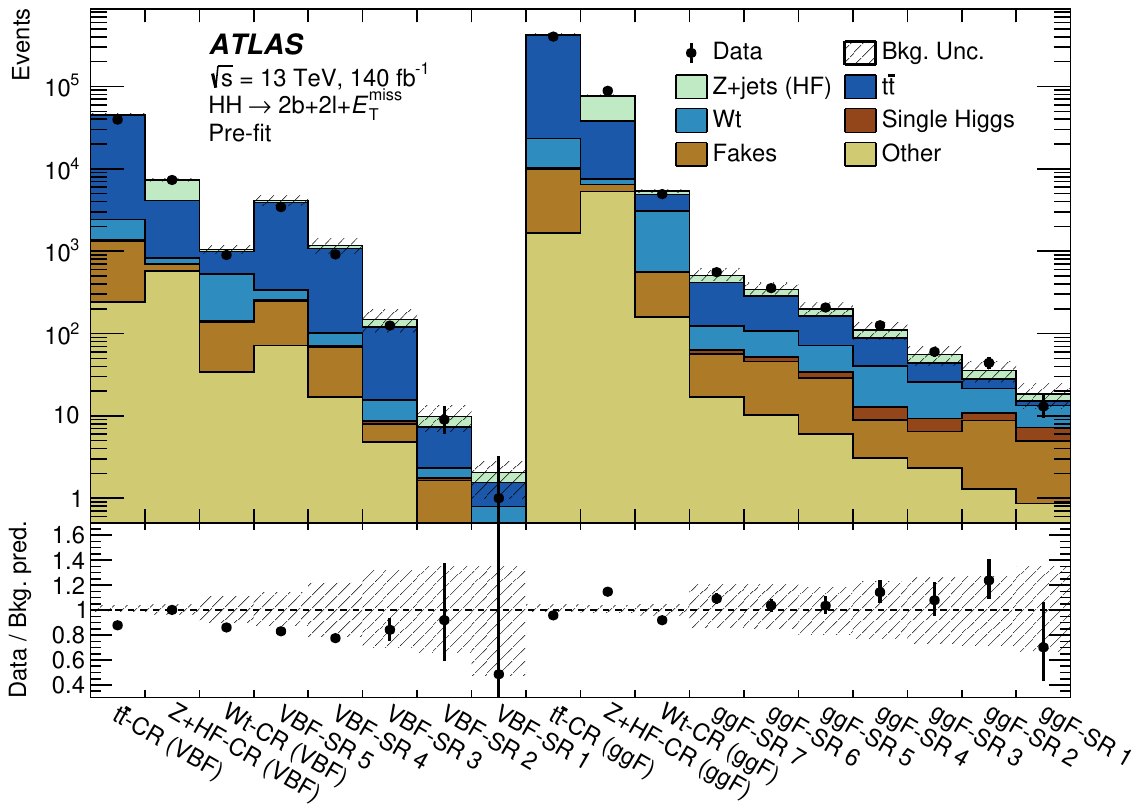}
\caption{Pre-fit yields of the $t\bar{t}$, $Z$+HF and $Wt$ CRs, both for the ggF and VBF event selection,
as well as the highest-score bins, numbered from high (VBF-SR~1 and ggF-SR~1) to low score
(VBF-SR~5 and ggF-SR~7), of the BDT and DNN output distribution in
the VBF and ggF event categories, respectively, as used in the final result.
The shaded bands include both statistical and systematic uncertainties.}
\label{fig:regions_prefit}
\end{figure}

The results account for several sources of systematic uncertainty on the signal and background processes,
which are classified as either experimental (detector- or luminosity-related) or theoretical modelling
uncertainties.
Statistical uncertainties of the simulated event samples are also taken into account.
The total pre-fit event yields, with their uncertainties, and the background composition in the different
signal and control regions, are shown in \cref{fig:regions_prefit}.
 
The uncertainties due to experimental sources are primarily due to the mismeasurement of reconstructed
object momentum and to the finite level of precision when determining reconstruction efficiencies.
They include uncertainties on the jet energy scale~\cite{PERF-2016-04} and jet energy resolution~\cite{ATL-PHYS-PUB-2015-015}.
Additional uncertainties for $b$-tagged jets arise from the precision of the $b$-tagging efficiency and
from the rates at which charm- and light-flavoured jets are selected as $b$-tagged jets~\cite{FTAG-2019-07}.
Lepton-related uncertainties arise on the electron~\cite{EGAM-2018-01} and muon~\cite{PERF-2014-07}
reconstructed energy (momentum) measurements, as well as on the precision of their reconstruction and identification efficiencies.
The \MET scale and resolution~\cite{PERF-2016-07} uncertainties, as well as uncertainties from the
mis-modelling of pile-up, trigger efficiency and luminosity, are also taken into account.
The uncertainty in the combined 2015--2018 integrated luminosity is \SI{0.83}{\percent}
\cite{DAPR-2021-01}, obtained using the LUCID-2 detector \cite{LUCID2} for the primary luminosity
measurements, complemented by measurements using the inner detector and calorimeters.
 
The normalization corrections on the dominant background processes, namely \ttbar, \wt and \zjets, are
determined using data in the control regions during the statistical analysis.
The experimental uncertainties and the systematic uncertainties from the modelling of these processes are
taken into account and constrained during the fitting process.
For the \ttbar and \wt processes, the QCD uncertainty is estimated by comparing events with different renormalization scale ($\mu_{R}$) and factorization scale ($\mu_{F}$) settings, where the largest deviation is chosen as the systematic uncertainty.
The uncertainties arising from the modelling of initial- and final-state radiation in the generators used
to simulate the \ttbar (\wt) background processes are evaluated using the method described in ref. \cite{ATL-PHYS-PUB-2016-004}.
The effects of uncertainties due to the choice of PDF and the value of $\alpha_\mathrm{S}$ are evaluated by varying the PDF
as well as the value for $\alpha_\mathrm{S}$ and taking the maximum variation into account as uncertainty.
The ME and PS uncertainties are estimated by comparing events from the nominal simulation samples with events from samples using alternative ME generators and PS generators, taking their difference as the respective uncertainty and then symmetrising it.
The uncertainty arising from the interference between the NLO predictions for \ttbar and \wt processes is estimated by taking the difference between the predicted yields obtained with the DR and DS schemes \cite{ATL-PHYS-PUB-2016-004}.
The \zjets modelling uncertainties are estimated using the nominal Sherpa 2.2.1 samples by considering
different merging and re-summation scales \cite{Hoeche:2009rj, Catani:2001cc}.
The uncertainties due to PDF variations and changes in $\mu_{R}$ and $\mu_{F}$ are calculated using the same procedures as for the \ttbar and \wt backgrounds.
 
Systematic uncertainties in the signal acceptance obtained by varying $\mu_{R}$ and $\mu_{F}$, as well as PDF-induced uncertainties, are taken into account as recommended~\cite{Grazzini:2018bsd} and evaluated using the same procedure as for the top-quark background processes.
The uncertainty due to the parton shower modelling is computed by comparing \Herwig[7] with \PYTHIA[8].
The uncertainty in the $HH$ production cross-section, evaluated to be $\pm\SI{3}{\percent}$ for
PDF+$\alpha_\mathrm{S}$ and ${}_{-\SI{23}{\percent}}^{+\SI{6}{\percent}}$ for the combined scale and top-quark mass
scheme for ggF, and $\pm\SI{2.1}{\percent}$ for
PDF+$\alpha_\mathrm{S}$ and $^{+\SI{0.03}{\percent}}_{-\SI{0.04}{\percent}}$ for the scale for VBF, is
included as an uncertainty in $\sigma^{SM}_{gg/qq\rightarrow HH/HHjj}$ when computing the upper limits on the signal strength.


\section{Statistical treatment and results}
\label{sec:result}

\subsection{Statistical procedure}\label{sec:result:statistical_procedure}
The statistical procedure used to interpret the data is described in \Refn{\cite{HIGG-2012-17}}.
The likelihood function is constructed from the product of Poisson probabilities:
\begin{equation*}
L (\text{data} | \mu, \bm{\theta})= \prod_{i=1}^N \text{Poisson}(\text{data}_i | \mu \cdot s_i(\bm{\theta}) + \mu_{b} b_i(\bm{\theta}))\times {\text G}(\tilde{\bm{\theta}} |\bm{\theta})
\end{equation*}
where $s_i$ and $b_i$ are the signal and background contributions in the $i$-th bin of the fitted
variable distribution; $\mu$ is the signal strength, $\mu_{b}$ is the normalization factor for the
respective background, and $\bm{\theta}$ denotes the nuisance parameters, which account for the uncertainties
of the measurements; ${\text G}(\tilde{\bm{\theta}} |\bm{\theta})$ is the Gaussian scaling function of the nuisance
parameters constructed as deviations from the nominal model of the systematic uncertainties, where
$\tilde{\bm{\theta}}$ provides a maximum likelihood estimate for $\bm{\theta}$. The parameter of interest in the
statistical analysis is the global signal strength factor $\mu = \sigma / \sigma^\mathrm{SM}$, which acts
on the total number of events predicted by the signal model. This factor is defined such that $\mu = 0$
corresponds to the background-only hypothesis and $\mu > 0 $ corresponds to a $HH$ signal in addition to
the background.
Hypothesised values of $\mu$ are tested based on the profile likelihood ratio~\cite{Cowan:2010js}, which compares data with background-only ($b$) and signal+background ($s+b$) models using the following test statistic:
\begin{equation*}
q_\mu =
\begin{cases}
-2 \text{ln} \frac{L( \text{data} | \mu, \hat{\hat{\bm{\theta}}}_\mu)}{L(\text{data} | 0, \hat{\bm{\theta}}_0)} & \hat{\mu} < 0 \\
-2 \text{ln} \frac{L( \text{data} | \mu, \hat{\hat{\bm{\theta}}}_\mu)}{L(\text{data} | \hat\mu, \hat{\bm{\theta}})} & 0 \leq \hat{\mu} \leq \mu \\
0 & \hat{\mu} > \mu    \\
\end{cases}
\end{equation*}
where $\hat\mu~\text{and}~\hat{\bm{\theta}}$ are the values
for $\mu$ and $\bm{\theta}$ when maximizing $L$ with all parameters floating (referred to as the unconditional
maximum-likelihood (ML) estimators). The $\hat{\hat{\theta}}_\mu$ is the conditional ML estimator of
$\theta$ for a fixed value of $\mu$.
This test statistic extracts the information on the signal strength from a full likelihood fit to the data.
The likelihood function includes all the parameters that describe the systematic uncertainties and their
correlations. The constraints on the coupling modifiers are obtained by using the respective modifier
($\kl$ or $\kvv$) in place of the signal strength as the parameter of interest in the profile likelihood
ratio.
 
Exclusion limits at $\SI{95}{\percent}$ CL are based on the $\mathrm{CL}_s$
prescription~\cite{Read:2002hq}, and they are set on the $HH$ production cross-section times branching
fraction divided by the corresponding SM prediction. The statistical analysis uses the distributions of
the MVA output score in both the ggF and VBF SRs as final discriminant.
The background CRs are used to constrain the overall normalization for the $\ttbar$, $Wt$ and $Z$+HF backgrounds in the SR.
Other background normalizations and shapes are fixed with prior uncertainties included in the fit.
 
\subsection{Limits on $HH$ production}
\begin{figure}
\centering
\includegraphics[width=0.8\textwidth]{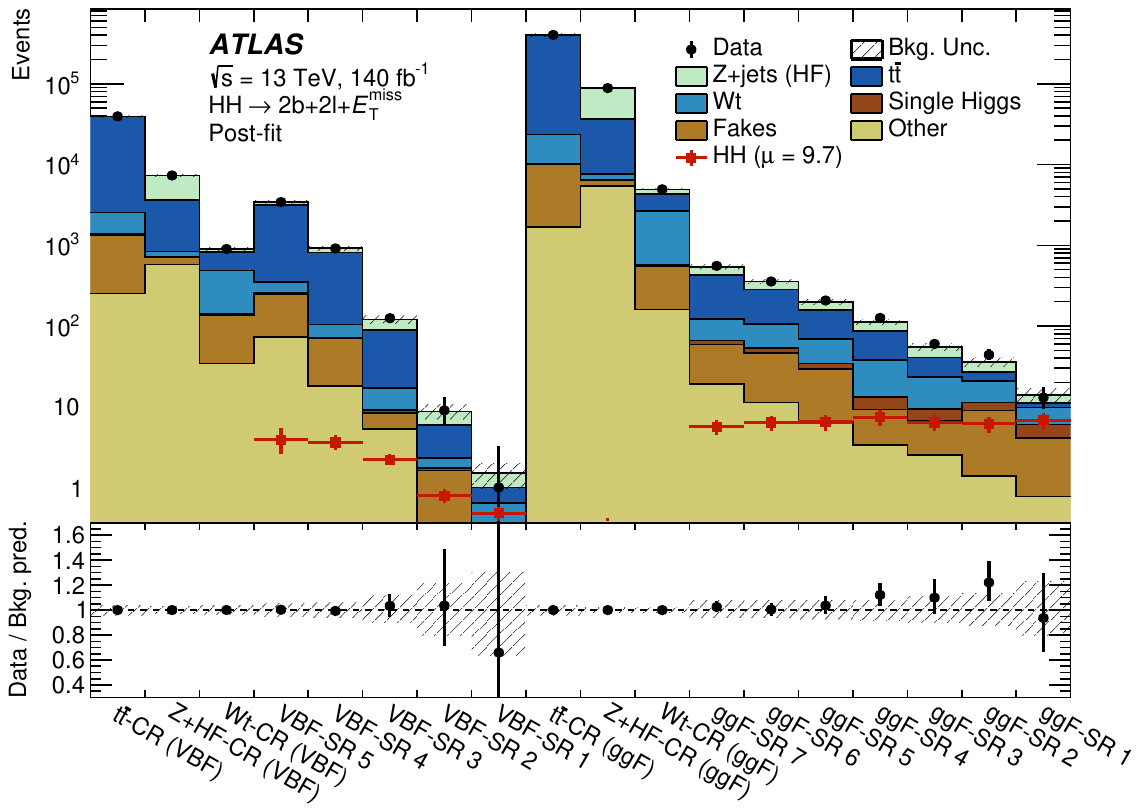}
\caption{Post-fit yields from the signal+background fit of the $t\bar{t}$, $Z$+HF and $Wt$ CRs,
both for the ggF and VBF event selections, as well as the highest-score bins, numbered from high
(VBF-SR~1 and ggF-SR~1) to low score (VBF-SR~5 and ggF-SR~7), of the BDT
and DNN output distribution in the VBF and ggF event categories respectively as used in
the final result. The fit is a conditional fit with the signal strength fixed to the
observed upper limit of $\mu_{HH} = 9.7$. The shaded bands include both statistical and
systematic uncertainties.}
\label{fig:result:nonres_2b:nn_output}
\end{figure}

\begin{figure}
\centering
\includegraphics[width=.7\textwidth]{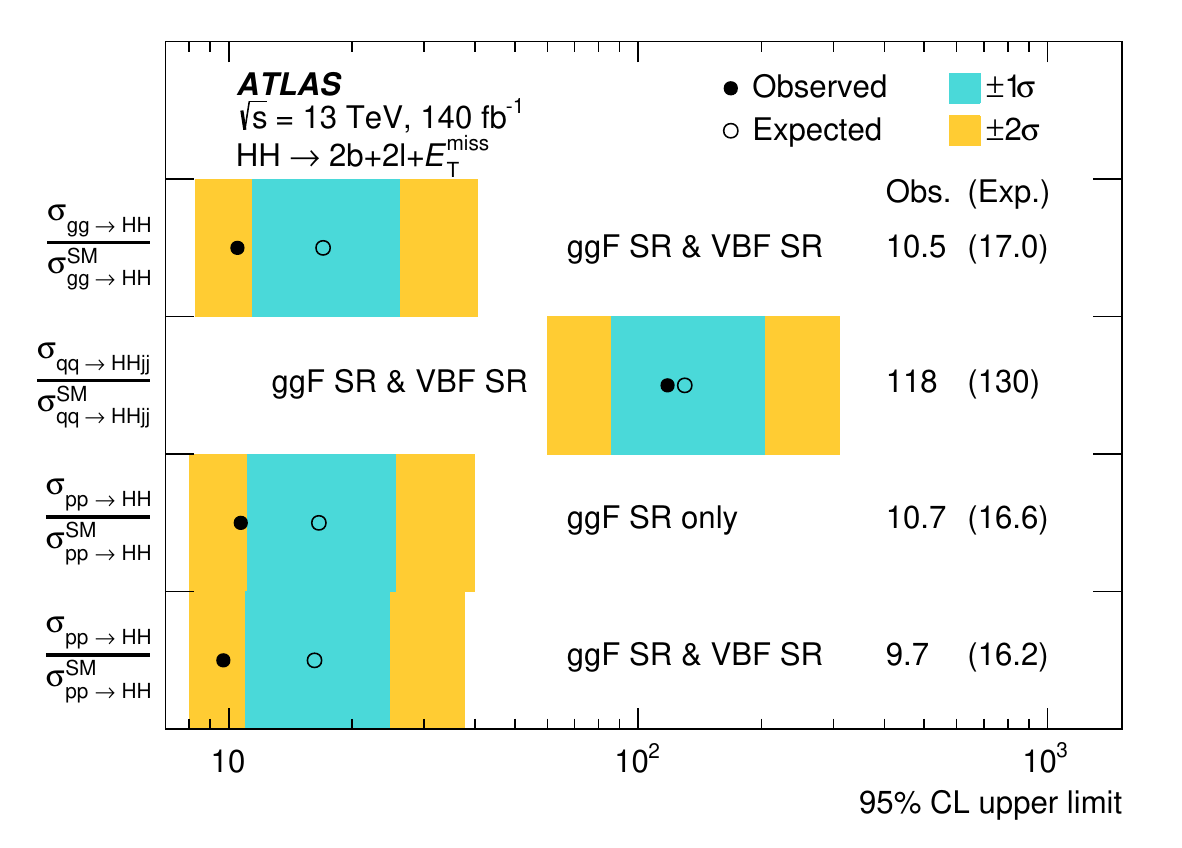}
\caption{Observed and expected upper limits on the ratios of the Higgs boson pair production
cross-section to the corresponding Standard Model prediction
$\sigma_{HH}/\sigma^\mathrm{SM}_{HH}$ for the ggF $HH$ signal only (top row), the VBF $HH$ signal only
while considering ggF $HH$ as background (second row) and the combined ggF+VBF $HH$ signal
considering only the ggF SR (third row) and considering all SRs (bottom row) at a
\SI{95}{\percent} confidence level. The relative ratio between the ggF and VBF production modes
is fixed to the SM value.}
\label{fig:result:nonres_2b:xsec_limits_syst}
\end{figure}

All regions used for this search are displayed in
\cref{fig:result:nonres_2b:nn_output} after a fit with the signal strength fixed to the upper limit, while pre-fit distributions are shown in
\cref{fig:regions_prefit}. A downward fluctuation of the data in the last bin of both the NN and BDT distributions is observed. A negative signal strength has been extracted from the fit: $\mu_{HH} = -8.5^{+7.7}_{-8.4}$.
Using the approach described in \cref{sec:result:statistical_procedure}, the upper limits on
the signal strength parameter for Higgs boson pair production with consideration of both the ggF and VBF signals are determined.
\Cref{fig:result:nonres_2b:xsec_limits_syst} summarizes the ggF, VBF
and combined results with the impact of all systematic uncertainties being shown. The observed combined upper
limit at \SI{95}{\percent} CL on the signal strength is $\mu_{HH} = 9.7$. In the CRs, the normalization and
modelling of the backgrounds play a prominent role, and they dominate the sources of uncertainty. In the
SRs, the systematic uncertainties mostly arise from background modelling, experimental sources and the
signal normalization. In the most sensitive bins, i.\,e. ggF-SR 1 to ggF-SR 3 and VBF-SR 1 and VBF-SR 2,
the statistical uncertainty becomes dominant due to the limited number of expected events.
 
\subsection{Constraints on Higgs coupling parameters}
\begin{figure}
\centering
\subfloat[][]{
\label{fig:nllscan:kl}
\includegraphics[width=0.45\textwidth]{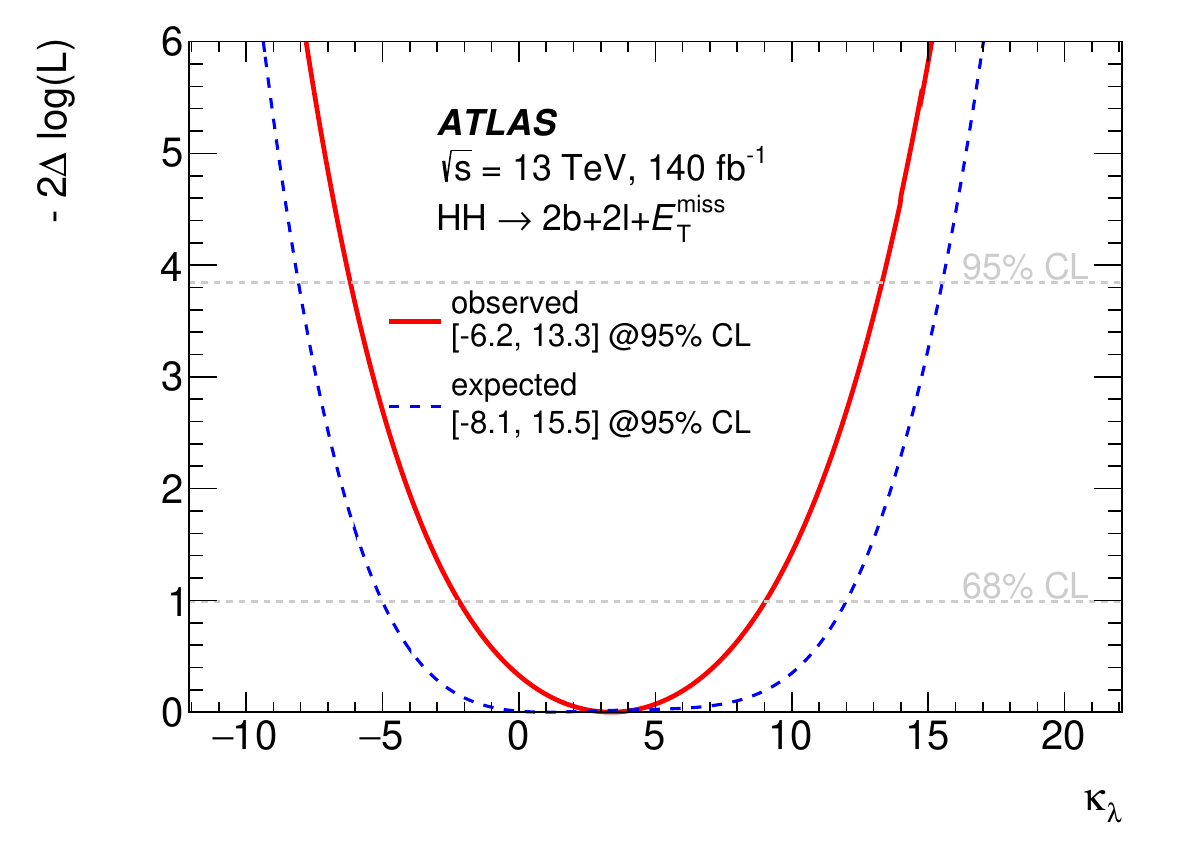}
}
\subfloat[][]{
\label{fig:nllscan:k2v}
\includegraphics[width=0.45\textwidth]{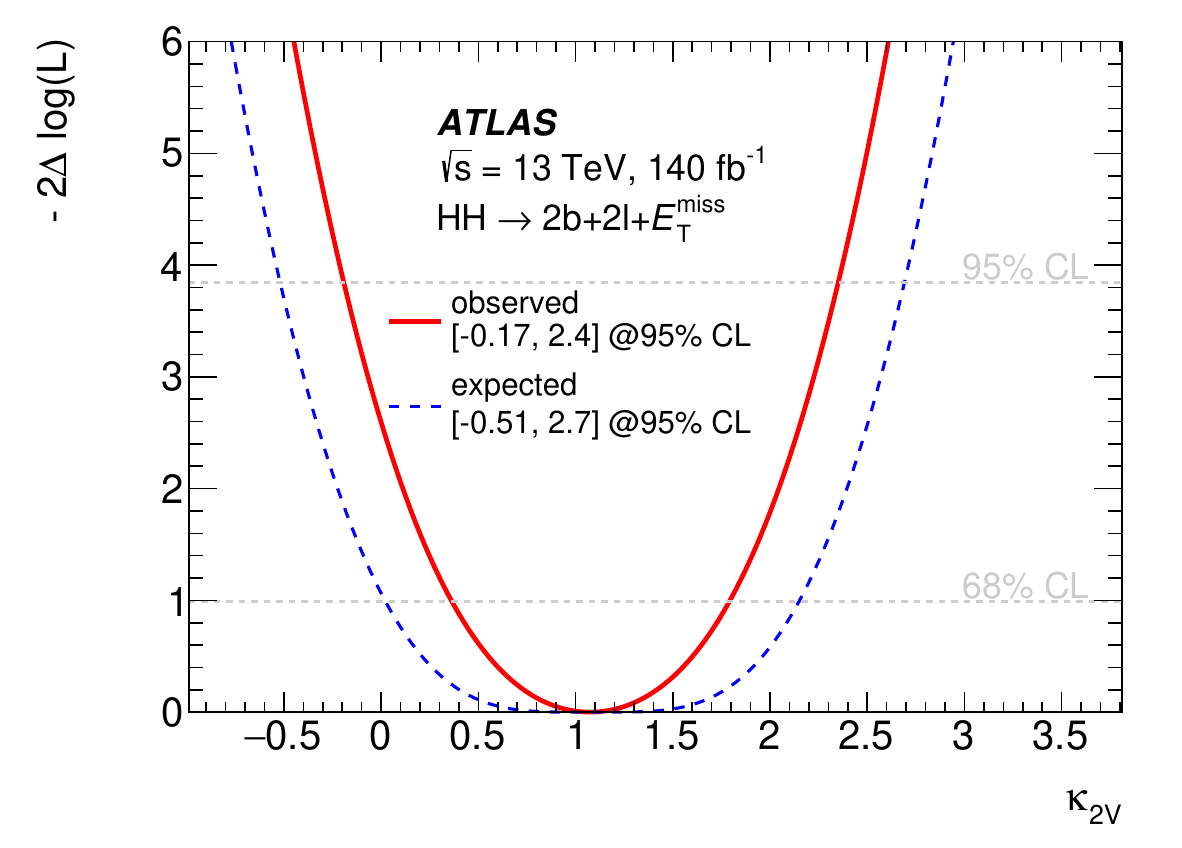}
}
\caption{Likelihood profiles of the (a) $\kl$ and (b) $\kvv$ parameters.}
\end{figure}
This analysis is extended by performing likelihood scans on the $\kl$ and $\kvv$ parameters.
The single Higgs boson background has a small dependence upon $\kl$ through loop effects, which is
neglected. Coupling modifiers other than the one tested in the respective scan are set to their
SM value. Hence, the $\kappa_t$, $\kappa_V$ and $\kvv$ coupling modifiers are set to $\kappa_t = \kappa_V
= \kvv = 1$ for the $\kl$-scan.
Both the ggF and VBF $HH$ signal regions are used in the analysis, and the result is shown in
\cref{fig:nllscan:kl}.
The observed result constrains $\kl$ to be within the range $[-6.2, 13.3]$ at \SI{95}{\percent} CL
which is slightly better than the expected range of $[-8.1, 15.5]$ at \SI{95}{\percent} CL due
to the observed downward fluctuation of the data.
 
A scan over the $\kvv$ parameter is also conducted. Again, all other couplings are set to their
respective SM values. Although only the VBF production mode of Higgs
boson pairs is sensitive to the $\kvv$ coupling modifier, both the ggF and VBF SRs are included in the
scan, allowing the analysis to achieve a slight enhancement of sensitivity from the presence of VBF $HH$ events within the ggF SR.
A likelihood scan is performed to set constraints on the $\kvv$ parameter; this is shown in
\cref{fig:nllscan:k2v}. Values for $\kvv$ are constrained to be within the range of $[-0.17, 2.4]$ at \SI{95}{\percent} CL
which is slightly better than the expected range of $[-0.51, 2.7]$ at \SI{95}{\percent} CL due
to the observed downward fluctuation of the data.


\FloatBarrier
 
\section{Conclusion}

A search for non-resonant Higgs boson pair production via the ggF and VBF production modes is performed.
It probes decay channels with one of the Higgs bosons decaying to $b\bar{b}$ and the other to either \WWstar,
\ZZstar, or \tautau. Selected events contain exactly two $b$-tagged jets, two light leptons with
opposite electric charge and missing transverse energy.
The analysis employs \SI{140}{\ifb} of $pp$ collision data at $\sqrt{s} = \SI{13}{TeV}$, recorded by the ATLAS detector at the LHC.
The results are consistent with the predictions for the SM background processes.
An observed (expected) \SI{95}{\percent} CL upper limit on the cross-section for the production of Higgs
boson pairs is set at 9.7 (16.2) times the SM prediction, which is a significant improvement compared to
the previous ATLAS search in this channel.
The analysis also establishes separate \SI{95}{\percent} CL limits on the Higgs coupling parameters \kl and \kvv excluding values outside the
ranges $[-6.2, 13.3]$ and $[-0.17, 2.4]$, respectively. These ranges are obtained
under the assumption that all other couplings, except the individual coupling being tested, are set to their SM values.


\section*{Acknowledgements}


We thank CERN for the very successful operation of the LHC, as well as the
support staff from our institutions without whom ATLAS could not be
operated efficiently.
 
We acknowledge the support of
ANPCyT, Argentina;
YerPhI, Armenia;
ARC, Australia;
BMWFW and FWF, Austria;
ANAS, Azerbaijan;
CNPq and FAPESP, Brazil;
NSERC, NRC and CFI, Canada;
CERN;
ANID, Chile;
CAS, MOST and NSFC, China;
Minciencias, Colombia;
MEYS CR, Czech Republic;
DNRF and DNSRC, Denmark;
IN2P3-CNRS and CEA-DRF/IRFU, France;
SRNSFG, Georgia;
BMBF, HGF and MPG, Germany;
GSRI, Greece;
RGC and Hong Kong SAR, China;
ISF and Benoziyo Center, Israel;
INFN, Italy;
MEXT and JSPS, Japan;
CNRST, Morocco;
NWO, Netherlands;
RCN, Norway;
MEiN, Poland;
FCT, Portugal;
MNE/IFA, Romania;
MESTD, Serbia;
MSSR, Slovakia;
ARRS and MIZ\v{S}, Slovenia;
DSI/NRF, South Africa;
MICINN, Spain;
SRC and Wallenberg Foundation, Sweden;
SERI, SNSF and Cantons of Bern and Geneva, Switzerland;
MOST, Taiwan;
TENMAK, T\"urkiye;
STFC, United Kingdom;
DOE and NSF, United States of America.
In addition, individual groups and members have received support from
BCKDF, CANARIE, Compute Canada and CRC, Canada;
PRIMUS 21/SCI/017 and UNCE SCI/013, Czech Republic;
COST, ERC, ERDF, Horizon 2020, ICSC-NextGenerationEU and Marie Sk{\l}odowska-Curie Actions, European Union;
Investissements d'Avenir Labex, Investissements d'Avenir Idex and ANR, France;
DFG and AvH Foundation, Germany;
Herakleitos, Thales and Aristeia programmes co-financed by EU-ESF and the Greek NSRF, Greece;
BSF-NSF and MINERVA, Israel;
Norwegian Financial Mechanism 2014-2021, Norway;
NCN and NAWA, Poland;
La Caixa Banking Foundation, CERCA Programme Generalitat de Catalunya and PROMETEO and GenT Programmes Generalitat Valenciana, Spain;
G\"{o}ran Gustafssons Stiftelse, Sweden;
The Royal Society and Leverhulme Trust, United Kingdom.
 
The crucial computing support from all WLCG partners is acknowledged gratefully, in particular from CERN, the ATLAS Tier-1 facilities at TRIUMF (Canada), NDGF (Denmark, Norway, Sweden), CC-IN2P3 (France), KIT/GridKA (Germany), INFN-CNAF (Italy), NL-T1 (Netherlands), PIC (Spain), ASGC (Taiwan), RAL (UK) and BNL (USA), the Tier-2 facilities worldwide and large non-WLCG resource providers. Major contributors of computing resources are listed in Ref.~\cite{ATL-SOFT-PUB-2023-001}.


\clearpage

\printbibliography
 
\clearpage
\input{atlas_authlist}

\end{document}

%% file: atlas_authlist.tex
 
\begin{flushleft}
\hypersetup{urlcolor=black}
{\Large The ATLAS Collaboration}

\bigskip

\AtlasOrcid[0000-0002-6665-4934]{G.~Aad}$^\textrm{\scriptsize 102}$,
\AtlasOrcid[0000-0002-5888-2734]{B.~Abbott}$^\textrm{\scriptsize 120}$,
\AtlasOrcid[0000-0002-1002-1652]{K.~Abeling}$^\textrm{\scriptsize 55}$,
\AtlasOrcid[0000-0001-5763-2760]{N.J.~Abicht}$^\textrm{\scriptsize 49}$,
\AtlasOrcid[0000-0002-8496-9294]{S.H.~Abidi}$^\textrm{\scriptsize 29}$,
\AtlasOrcid[0000-0002-9987-2292]{A.~Aboulhorma}$^\textrm{\scriptsize 35e}$,
\AtlasOrcid[0000-0001-5329-6640]{H.~Abramowicz}$^\textrm{\scriptsize 151}$,
\AtlasOrcid[0000-0002-1599-2896]{H.~Abreu}$^\textrm{\scriptsize 150}$,
\AtlasOrcid[0000-0003-0403-3697]{Y.~Abulaiti}$^\textrm{\scriptsize 117}$,
\AtlasOrcid[0000-0002-8588-9157]{B.S.~Acharya}$^\textrm{\scriptsize 69a,69b,m}$
% ...
\bigskip
\\

% ...
$^{29}$Physics Department, Brookhaven National Laboratory, Upton NY; United States of America.\\
% ...
$^{35}$$^{(a)}$Facult\'e des Sciences Ain Chock, R\'eseau Universitaire de Physique des Hautes Energies - Universit\'e Hassan II, Casablanca;$^{(b)}$Facult\'{e} des Sciences, Universit\'{e} Ibn-Tofail, K\'{e}nitra;$^{(c)}$Facult\'e des Sciences Semlalia, Universit\'e Cadi Ayyad, LPHEA-Marrakech;$^{(d)}$LPMR, Facult\'e des Sciences, Universit\'e Mohamed Premier, Oujda;$^{(e)}$Facult\'e des sciences, Universit\'e Mohammed V, Rabat;$^{(f)}$Institute of Applied Physics, Mohammed VI Polytechnic University, Ben Guerir; Morocco.\\
% ...
$^{49}$Fakult\"{a}t Physik , Technische Universit{\"a}t Dortmund, Dortmund; Germany.\\
% ...
$^{55}$II. Physikalisches Institut, Georg-August-Universit\"{a}t G\"ottingen, G\"ottingen; Germany.\\
% ...
$^{69}$$^{(a)}$INFN Gruppo Collegato di Udine, Sezione di Trieste, Udine;$^{(b)}$ICTP, Trieste;$^{(c)}$Dipartimento Politecnico di Ingegneria e Architettura, Universit\`a di Udine, Udine; Italy.\\
% ...
$^{102}$CPPM, Aix-Marseille Universit\'e, CNRS/IN2P3, Marseille; France.\\
% ...
$^{117}$Department of Physics, New York University, New York NY; United States of America.\\
% ...
$^{120}$Homer L. Dodge Department of Physics and Astronomy, University of Oklahoma, Norman OK; United States of America.\\
% ...
$^{150}$Department of Physics, Technion, Israel Institute of Technology, Haifa; Israel.\\
$^{151}$Raymond and Beverly Sackler School of Physics and Astronomy, Tel Aviv University, Tel Aviv; Israel.\\
% ...
$^{a}$ Also Affiliated with an institute covered by a cooperation agreement with CERN.\\
$^{b}$ Also at An-Najah National University, Nablus; Palestine.\\
$^{c}$ Also at Borough of Manhattan Community College, City University of New York, New York NY; United States of America.\\
$^{d}$ Also at Center for High Energy Physics, Peking University; China.\\
$^{e}$ Also at Center for Interdisciplinary Research and Innovation (CIRI-AUTH), Thessaloniki; Greece.\\
$^{f}$ Also at Centro Studi e Ricerche Enrico Fermi; Italy.\\
% ...
$^{m}$ Also at Department of Physics, King's College London, London; United Kingdom.\\
% ...

\end{flushleft}
